\documentclass[12pt,draftcls,onecolumn]{IEEEtran}
\ifCLASSINFOpdf
\else
\fi
\usepackage[cmex10]{amsmath}
\usepackage{amssymb}
\usepackage{mathrsfs}
\usepackage{verbatim}
\usepackage{graphicx}
\usepackage{epstopdf}
\usepackage{cite}
\usepackage[justification=justified]{caption}
\usepackage{ifthen}

\newboolean{singleColumn}
\setboolean{singleColumn}{true}

\DeclareMathOperator{\sinc}{sinc} 
\DeclareMathOperator{\rect}{rect}

\DeclareMathOperator{\PMEPR}{PMEPR}

\begin{document}
\title{Adaptive Transmit Waveform Design using \\ Multi-Tone Sinusoidal Frequency Modulation}

\author{David~A.~Hague,~\IEEEmembership{Member,~IEEE,}}% <-this % stops a space
\maketitle

\begin{abstract}
This paper presents an adaptive waveform design method using Multi-Tone Sinusoidal Frequency Modulation (MTSFM). The MTSFM waveform's modulation function is represented as a finite Fourier series expansion.  The Fourier coefficients are utilized as a discrete set of design parameters that may be modified to adapt the waveform's properties.  The MTSFM's design parameters are adjusted to shape the spectrum, Auto-Correlation Function (ACF), and Ambiguity Function (AF) shapes of the waveform.  The MTSFM waveform model naturally possesses the constant envelope and spectrally compact waveforms that make it well suited for transmission on practical radar/sonar transmitters which utilize high power amplifiers.  The MTSFM has an exact mathematical definition for its time-series using Generalized Bessel Functions which allow for deriving closed-form analytical expressions for its spectrum, AF, and ACF.  These expressions allow for establishing well-defined optimization problems that finely tune the MTSFM's properties.  This adaptive waveform design model is demonstrated by optimizing MTSFM waveforms that initially possess a ``thumbtack-like'' AF shape.  The resulting optimized designs possess substantially improved sidelobe levels over specified regions in the range-Doppler plane without increasing the Time-Bandwidth Product (TBP) that the initialized waveforms possessed.  Simulations additionally demonstrate that the optimized thumbtack-like MTSFM waveforms are competitive with thumbtack-like phase-coded waveforms derived from design algorithms available in the published literature.  
\end{abstract}


% \fi
\IEEEpeerreviewmaketitle
\section{Introduction}

%% ADD this into first intro paragraph later
%One of the components of a cognitive system that perhaps possesses the greatest design adaptability is the transmit waveform. 

\IEEEPARstart{W}{aveform} diversity has been a topic of great interest, particularly in the radar community, for the last two decades \cite{BluntIV}.  The growth experienced in this field has been inspired by the preeminence of cognitive systems which seek to leverage information gathered from earlier interactions with the environment to inform the selection of system parameters to optimize system performance \cite{HaykinI, CognitiveI}.  Cognitive systems exploit parameterized waveform models that enable synthesizing a diverse set of  waveforms with unique properties.  There are a number of waveform properties that may be adjusted to optimize system performance including the waveform's operational band of frequencies, pulse-length, and transmit power to name a few.  A set of properties collectively referred to as waveform shape are of particular interest.  Waveform shape refers to either the time-frequency characteristics of the waveform's modulation function which in turn informs its overall spectral shape as well as the shape of its Ambiguity Function (AF) and its zero Doppler counterpart, the Auto Correlation Function (ACF).  These metrics for waveform shape are often utilized due to their foundational applicability to many practical systems and further reinforced by the rigorous mathematical results that exist to describe their structure \cite{Cook, Rihaczek, Levanon}.

The ability to adapt waveform shape requires a parameterized waveform model that ideally facilitates adaptation according to some set of well defined optimality metrics.  Since Woodward's seminal work which introduced the AF \cite{Woodward}, there has been a wealth of research focusing on the problem of optimizing a waveform to possess certain waveform shape properties, see \cite{Wilcox, Sussman, GladkovaI, GladkovaII} for an overview of the fundamentals regarding these techniques.  The vast majority of waveform shape design research has focused on developing a wide variety of algorithms to generate Phase-Coded (PC) waveforms \cite{JianLiBookII}.  Theoretically there exists a nearly endless combination of phase codes that can be employed making PC waveforms an extremely versatile parameterized waveform model.  There continues to be extensive research on designing optimal PC waveforms for Multiple-Input Multiple-Output (MIMO) applications \cite{jianLiII, PalomarIII, RangaswamyI} and cognitive radar applications \cite{AubryII, AubryIII, PrabhuBabuII}.  Additionally, the general study of developing algorithms to design PC waveforms with specific ACF/AF properties is still a problem of interest to the radar and sonar communities \cite{AubryI, PrabhuBabuIII, PalomarI, PalomarII, SoltanalianI}.

In addition to waveform shape, there are a number of design issues to consider when transmitting waveforms on practical systems.  It is generally desirable for a waveform to possess a constant envelope which translates to having a low Peak-to-Mean Envelope Power Ratio (PMEPR).  This is required to reduce the distortion that amplitude modulation introduces to a saturated power amplifier, a common electronic component in most radar/sonar transmitters.  Another challenge is to design a waveform whose energy resides in a compact band of operational frequencies with little energy residing outside of that band.  This is an important property as most practical transmitters either have a limited bandwidth which filters out of band energy or have a frequency response that is not an ideal all-pass system.  Transmitted waveforms with substantial spectral extent on such systems will distort the resulting signal that is transmitted into the medium and risks degrading their waveform shape properties.  This spectral compactness property is referred to as Spectral Efficiency (SE) and there exist explicit mathematical definitions to measure this waveform property \cite{BluntIV, HagueIV}.  High SE is most readily accomplished by a waveform whose phase/frequency modulation functions are smooth and do not contain any abrupt transitions in instantaneous phase or frequency.  

Most Frequency Modulated (FM) waveforms naturally possess both a constant envelope (i.e, a low PMEPR) and high SE making them well suited for transmission on practical devices.  However, most FM waveform models possess very few design parameters that allow for adapting waveform shape which places strict limits on their design versatility.  While PC waveforms possess tremendous design versatility and are generally constant envelope, they do not naturally possess high SE.  PC waveforms have substantial spectral extent due to the transient-like phase transitions between chips \cite{Levanon}. This has motivated the development of Continuous Phase Modulation (CPM) techniques to improve upon their spectral characteristics \cite{Blinchikoff, BluntI, BluntIII} by introducing continuity in the first few derivatives of the waveform's instantaneous phase.  These CPM methods must also deal with minimizing the distortion of the waveform's AF shape \cite{LevanonQuad, BluntII} that naturally arises from modifying the waveform's instantaneous phase.  Nevertheless, the design versatility of parameterized waveform models is an attractive feature as long as the SE issues can be mitigated. 

The CPM methods aimed at improving the SE of PC waveforms \cite{BluntI, BluntIII} effectively transform PC waveforms into spectrally compact parameterized FM waveforms by introducing continuity in the first few derivatives of the waveform's instantaneous phase.  This combines the constant envelope and spectrally compact properties of FM waveforms while also introducing a discrete set of design parameters that PC waveforms possess.  Inspired by these paramterized FM waveform models, this paper describes a constant evenlope spectrally compact adaptive waveform model using Multi-Tone Sinusoidal Frequency Modulation (MTSFM). The MTSFM waveform's modulation function is represented as a finite sum of weighted sinusoidal functions expressed as a Fourier series expansion.  The Fourier coefficients are then utilized as a finite discrete set of design parameters.  These design parameters are then adjusted to modify waveform shape properties.  The MTSFM belongs to the family of general multi-carrier waveforms \cite{Koivunen, Levanon} and bears a strong resemblance to various Orthogonal Frequency Division Multiplexing (OFDM) and Constant-Envelope OFDM (CE-OFDM) techniques \cite{CEOFDM, Nehorai}.  Moreover, the MTSFM waveform's time-series can be expressed in a precise analytical form using Generalized Bessel Functions (GBF) \cite{Dattoli}.  This model allows for deriving exact closed form expressions that precisely describe the MTSFM's waveform shape properties.  These expressions aid in defining appropriate optimization problems that finely tune the MTSFM's properties enabling physically realizable adaptive waveforms.  This GBF-based mathematical representation is potentially applicable to the analysis and synthesis  of the waveform shape properties of other multi-carrier waveform models such OFDM and CE-OFDM.  The rest of this paper is organized as follows: Section \ref{sec:waveformModel} defines the waveform signal model.  Section \ref{sec:MTSFM} defines the MTSFM waveform model and demonstrates the model via illustrative design examples.  Section \ref{sec:MTSFM_Performance} more thoroughly evaluates the performance of the MTSFM and compares it to other established PC waveform design methods available in the published literature.  Lastly, Section \ref{sec:Conclusion} presents the conclusions of the paper.

\section{Transmit Waveform Signal Model and Measures of Performance}
\label{sec:waveformModel}
This section describes the waveform complex analytic signal model, AF and ACF.  This model assumes a mono-static radar/sonar system where the target of interest is a point target undergoing constant velocity motion. 
\subsection{The Complex Analytic Model} 
\label{subsec:complexAnalytic}
The transmit waveform signal $s\left(t\right)$ is modeled as a complex analytic signal with total energy $E$ and pulse-length $T$ defined over the interval $-T/2 \leq t \leq T/2$ expressed as
\begin{IEEEeqnarray}{rCl}
s\left(t\right) = a\left(t\right)e^{j\varphi\left(t\right)}e^{j2\pi f_c t}
\label{eq:complexExpo}
\end{IEEEeqnarray}  
where $\varphi\left(t\right)$ is the phase modulation function of the waveform, $f_c $ is the carrier frequency, and $a\left(t\right)$ is a real-valued and positive amplitude tapering function \cite{Rihaczek}.  For all the design examples in this paper, a Tukey window with shape parameter $\alpha_T$ \cite{Harris} will be utilized as the amplitude tapering function.  The shape parameter $\alpha_T$ allows for smoothly trading off between a rectangular window ($\alpha_T = 0.0$) and a Hann window ($\alpha_T = 1.0$).  Unless otherwise specified, the waveform model \eqref{eq:complexExpo} will utilize a shape parameter $\alpha_T = 0.0$ and assumes the waveform is basebanded (i.e, $f_c =0$).  The waveform model in \eqref{eq:complexExpo} then simplifies to 
\begin{IEEEeqnarray}{rCl}
s\left(t\right) = \dfrac{\rect\left(t/T\right)}{\sqrt{T}}e^{j\varphi\left(t\right)}
\label{eq:complexExpo1}
\end{IEEEeqnarray} 
where the $1/\sqrt{T}$ term normalizes the waveform to possess unit energy.  The model \eqref{eq:complexExpo1} will be used throughout the paper to derive closed form expressions for various performance measures of the MTSFM waveform model.  Additionally, the waveform that results from \eqref{eq:complexExpo1} has an instantaneous frequency function that does not possess any AM contributions and is therefore solely determined by its modulation function.  The waveform's modulation function is expressed as 
\begin{IEEEeqnarray}{rCl}
m\left(t\right) = \dfrac{1}{2 \pi}\dfrac{d\left[\varphi \left( t\right)\right]}{dt}.
\label{eq:m}
\end{IEEEeqnarray}  

The transmitter electronics of a radar or sonar system are generally peak power limited and operate efficiently when the transmit waveform possesses a constant envelope.  The degree to which a waveform's envelope is constant can be measured using the Peak to Mean Envelope Power Ratio (PMEPR) \cite{Levanon}.  The PMEPR is defined as the square of the Crest Factor (CF) expressed in dB as
\begin{IEEEeqnarray}{rCl}
\PMEPR = 10\log_{10}\Biggl\{\left(\dfrac{\max_t\{|s\left(t\right)|^2\}}{\frac{1}{T}\int_{-T/2}^{T/2}|s\left(t\right)|^2dt}\right)\Biggr\}
\label{eq:PMEPR}
\end{IEEEeqnarray}
The PMEPR provides a measure of the total energy of waveforms with the same duration $T$.  A low PMEPR translates to a high average power and therefore higher total energy.  Using a rectangular amplitude tapering function as in \eqref{eq:complexExpo1} results in a minimum PMEPR of 0 dB.  Any tapering of the waveform (i.e., increasing the Tukey window shape parameter $\alpha_T$) will increase its PMEPR from this optimal value resulting in a waveform with less total energy.  An additional requirement for a waveform to be well suited for transmission on practical electronics is for it to possess high SE.   One commonly utilized method of measuring SE that provides a fair means of comparison between waveforms is that of \cite{BluntIV, HagueIV} which defines the SE $\Theta\left(W\right)$ as the ratio of waveform energy in a specific band of frequencies $W$ centered on $f_c$ to the total energy of the waveform across all frequencies expressed as 
\begin{IEEEeqnarray}{rCl}
\Theta\left(W\right) = \dfrac{\int_{f_c-W/2}^{f_c+W/2}|S\left(f\right)|^2df}{\int_{-\infty}^{\infty}|S\left(f\right)|^2df} = \int_{-W/2}^{W/2}|S\left(f\right)|^2df.
\label{eq:THETA}
\end{IEEEeqnarray}
where $S\left(f\right)$ is the waveform's Fourier transform.  Note that the second integral results from the assumption that the waveform's energy in the denominator is unity and  the waveform is basebanded.  

\subsection{The Ambiguity Function} 
\label{subsec:ambiguityFunction}
This signal model assumes a Matched Filter (MF) receiver is used to process target echoes.  The MF, also known as a correlation receiver, is the optimal detection receiver for a known signal embedded additive white Gaussian noise \cite{Cook}.  The Ambiguity Function (AF) measures the response of the waveform's MF to its Doppler shifted versions and is defined as \cite{Rihaczek, Cook}
\begin{IEEEeqnarray}{rCl}
\chi\left(\tau, \nu\right) = \int_{-\infty}^{\infty}s\left(t-\frac{\tau}{2}\right)s^*\left(t+\frac{\tau}{2}\right)e^{j2\pi \nu t} dt
\label{eq:AF}
\end{IEEEeqnarray}
where $\nu$ is the doppler shift expressed as $\nu = \frac{2\dot{r}}{c}f_c$.  Note that the AF defined in \eqref{eq:AF} models the narrowband Doppler effect.  Unlike its broadband counterpart which represents the general Doppler scaling effect, this variant of the AF possess more convenient mathematical properties which simplifies the analysis of the MTSFM waveform design model.  Additionally, the narrowband approximation is generally accurate for most radar and many sonar system applications.  Lastly, the ACF is the zero Doppler cut of the AF
\ifthenelse {\boolean{singleColumn}}
 {\begin{IEEEeqnarray}{rCl}
R\left(\tau\right) = \chi\left(\tau, \nu\right)|_{\nu=0} = \int_{-\infty}^{\infty}s\left(t-\frac{\tau}{2}\right)s^*\left(t+\frac{\tau}{2}\right) dt
\label{eq:ACF}
\end{IEEEeqnarray}}
{\begin{multline}
R\left(\tau\right) = \chi\left(\tau, \nu\right)|_{\nu=0} \\ = \int_{-\infty}^{\infty}s\left(t-\frac{\tau}{2}\right)s^*\left(t+\frac{\tau}{2}\right) dt
\label{eq:ACF}
\end{multline}}
This paper, like most results in the published literature, will focus on the modulus squared of the AF $\left|\chi\left(\tau, \nu\right)\right|^2$ and ACF $\left|R\left(\tau\right)\right|^2$.  There exist explicit mathematical properties describing the distribution of the volume of $\left|\chi\left(\tau, \nu\right)\right|^2$ in the range-Doppler plane and a similar analysis can be performed on the modulus of the ACF $|R\left(\tau\right)|^2$.

Waveforms may possess a wide variety of AF shapes with mainlobe and sidelobe structure that is intimately linked with the time-frequency characteristics of the waveform's modulation function \cite{Cook, Rihaczek, Ricker}.  This paper will specifically focus on the design of waveforms that possess a thumbtack-like AF.  These waveforms attain an AF with a mainlobe whose width in range and Doppler is inversely proportional to the waveform's bandwidth and pulse-length respectively.  There is ideally zero or at worst non-zero but negligibly small coupling between the range and Doppler mainlobe structure.  This allows for resolving multiple targets distributed in the range-Doppler plane.  The rest of the AF's bounded volume is spread uniformly in the range-Doppler plane \cite{Cook, Rihaczek, Levanon} resulting in a pedestal of sidelobes whose height is inversely proportional to the waveform's Time-Bandwidth Product (TBP).  

The uncoupled mainlobe structure and uniform distribution of sidelobe levels of the thumbtack AF shape simplifies the analysis and comparison of various waveform design models and is one of the main reasons why this paper focuses on the design of thumbtack-like waveforms.  Optimizing a thumbtack-like waveform is of practical interest as well.  The TBP establishes the height of the pedastal of sidelobes that is evenly distributed in the range-Doppler plane.  For large TBP waveforms, the sidelobe levels may be acceptably low enough to distinguish a weak target in the presence of a much stronger one.  However, many systems are limited in how large a TBP waveform they can reliably generate.  This means the pedestal of sidelobes can become unacceptably high and weak targets get masked by echoes from stronger target returns.  Reducing the pedestal height of a thumbtack-like waveform's AF over sub-regions in the range-Doppler plane could help alleviate this issue for small TBP waveforms.    

\section{The Multi-Tone Sinsudoial Frequency Modulated Waveform Model}
\label{sec:MTSFM}
This section describes the MTSFM model and how it can be used to synthesize waveforms with desired AF/ACF shapes.  These techniques are then demonstrated via illustrative design examples.

\subsection{The MTSFM Waveform Model}
\label{subsec:MTSFM_Intro}
The MTSFM waveform is created by representing the modulation function \eqref{eq:m} as a Fourier series expansion. The modulation function is expressed in terms of even and odd symmetric harmonics as
\begin{align}
m\left(t\right) &= m_e\left(t\right) + m_o\left(t\right) \\ &= \frac{a_0}{2} + \sum_{k=1}^K a_k \cos\left(\frac{2 \pi k t}{T}\right) + b_k \sin\left(\frac{2 \pi k t}{T}\right).
\label{eq:MTSFM_1}
\end{align}
where $m_e\left(t\right)$ and $m_o\left(t\right)$ are respectively the even and odd symmetric components of the Fourier series expansion
\begin{align}
m_e\left(t\right) &= \frac{a_0}{2} + \sum_{k=1}^Ka_k \cos\left(\frac{2 \pi k t}{T}\right),\label{eq:memo1} \\
m_o\left(t\right)&= \sum_{k=1}^Kb_k \sin\left(\frac{2 \pi k t}{T}\right). 
\label{eq:memo2}
\end{align}
Integrating with respect to time and multiplying by $2\pi$ yields the phase modulation function of the waveform expressed as
\begin{align}
\varphi\left(t\right) &= \varphi_e\left(t\right)+ \varphi_o\left(t\right) \\ &= \pi a_0t + \sum_{k=1}^K \alpha_k \sin\left(\frac{2 \pi k t}{T}\right) - \beta_k \cos\left(\frac{2 \pi k t}{T}\right)
\label{eq:MTSFM_2}
\end{align}
where $\varphi_e\left(t\right)$ and $\varphi_o\left(t\right)$ are the instantaneous phase functions derived from the even and odd modulation functions \eqref{eq:memo1} and \eqref{eq:memo2}
\begin{align}
\varphi_e\left(t\right) &= \pi a_0t + \sum_{k=1}^K \alpha_k \sin\left(\frac{2 \pi k t}{T}\right), \\
\varphi_o\left(t\right) &= -\sum_{k=1}^K \beta_k \sin\left(\frac{2 \pi k t}{T}\right) 
\label{eq:MTSFM_3}
\end{align}
and $\{\alpha_k, \beta_k\}_{k=1}^K$ are the waveform's modulation indices expressed as $\Bigl\{\left(\frac{a_k T}{k}\right), \left(\frac{b_k T}{k}\right)\Bigr\}_{k=1}^K$.  This paper will simply denote the set of modulation indices $\{\alpha_k, \beta_k\}_{k=1}^K$ as $\{\alpha_k, \beta_k\}$.  The even/odd modulation and instantaneous phase functions are explicitly defined here because MTSFM waveforms with either even or odd symmetry in their modulation functions have distinct AF/ACF characteristics.  These properties will be demonstrated later in the paper.  The more general model \eqref{eq:MTSFM_2} blends these characteristics thus obscuring their unique symmetry properties.    Inserting \eqref{eq:MTSFM_2} into the basebanded version of the waveform signal model \eqref{eq:complexExpo1} yields the MTSFM waveform time-domain representation
\ifthenelse {\boolean{singleColumn}}
 {\begin{IEEEeqnarray}{rCl}
s\left(t\right) = \dfrac{\rect\left(t/T\right)}{\sqrt{T}}\exp\Biggl\{j\sum_{k=1}^K \alpha_k \sin\left(\frac{2 \pi k t}{T}\right) - \beta_k \cos\left(\frac{2 \pi k t}{T}\right) \Biggr\}.
\label{eq:MTSFM_4}
\end{IEEEeqnarray}}
{\begin{multline}
s\left(t\right) = \dfrac{\rect\left(t/T\right)}{\sqrt{T}} \times \\ \exp\Biggl\{j\sum_{k=1}^K \alpha_k \sin\left(\frac{2 \pi k t}{T}\right) - \beta_k \cos\left(\frac{2 \pi k t}{T}\right) \Biggr\}.
\label{eq:MTSFM_4}
\end{multline}}
This direct implementation of \eqref{eq:complexExpo1} results in an expression that does not readily allow for solving closed form expressions for waveform shape properties.  

However, the MTSFM can be represented in a manner that does permit closed-form expressions for waveform shape properties.  This is achieved by expressing \eqref{eq:MTSFM_4} as a complex Fourier series expansion
\begin{IEEEeqnarray}{rCl}
 s\left(t\right) =  \dfrac{\rect\left(t/T\right)}{\sqrt{T}}\sum_{m=-\infty}^{\infty} c_m  e^{j\frac{2\pi m t}{T}}e^{j\pi a_0 t}.
\label{eq:MTSFM_5}
\end{IEEEeqnarray}
The Fourier coefficients, as shown in Appendix \ref{sec:AppendixI}, can be expressed in exact closed form in terms of three types of GBFs depending on the symmetry of the waveform's modulation function
\begin{equation}  c_m = \left\{
\begin{array}{ll}
	\mathcal{J}_m^{1:K}\left(\{\alpha_k, -j\beta_k\}\right), & \varphi\left(t\right) \\

      \mathcal{J}_m^{1:K}\left(\{\alpha_k\}\right), & \varphi_e\left(t\right) \\
      
      \mathcal{I}_m^{1:K}\left(\{-j\beta_k\}\right), & \varphi_o\left(t\right) \\
\end{array} 
\right.
\label{eq:SFM_Fourier_Series} 
\end{equation}
where $\mathcal{J}_m^{1:K}\left(\{\alpha_k, -j\beta_k\}\right)$ is the $K$-dimensional GBF of the mixed-type, $\mathcal{J}_m^{1:K}\left(\{\alpha_k\}\right)$ is the cylindrical $K$-dimensional GBF, and $ \mathcal{I}_m^{1:K}\left(\{-j\beta_k\}\right)$ is the $K$-dimensional Modified GBF (M-GBF) \cite{Lorenzutta}.  The expression in \eqref{eq:MTSFM_5} represents the MTSFM in terms of Wilcox's model \cite{Wilcox} where the orthonormal basis functions are the complex exponentials $e^{j\frac{2\pi mt}{T}}$ and the Fourier coefficients $c_m$ are the $m^{th}$ order GBFs shown in \eqref{eq:SFM_Fourier_Series}.  This representation of the MTSFM now readily allows for deriving closed form expressions for a wide variety of performance metrics including the spectrum, AF, and ACF.  

The MTSFM waveform model naturally possesses a constant envelope \cite{HagueIV} which satisfies the first primary requirement for transmitting waveforms on practical electronics.  Additionally, the MTSFM's modulation function is expressed as a finite Fourier series.  Any finite Fourier series is continuous and infinitely differentiable \cite{boyd}.  Therefore the modulation function is smooth and does not contain any transient-like discontinuities unlike PC waveforms.  The smoothness of the MTSFM's modulation function would require several stages of CPM to approximate.  As a result of these smoothness properties, the vast majority of the MTSFM waveform's energy will be densely concentrated in its swept bandwidth $\Delta f$ with very little energy residing outside of that band.  The spectrum of the MTSFM waveform is expressed as \cite{HagueDiss, HagueIV}
\ifthenelse {\boolean{singleColumn}}
 {\begin{IEEEeqnarray}{rCl}
S\left(f\right) = \sqrt{T} \sum_{m=-\infty}^{\infty}\mathcal{J}_m^{1:K}\left(\{\alpha_k,-j\beta_k\}\right)\sinc\left[\pi T \left(f-\frac{m}{T}\right)\right].
\label{eq:MTSFM_Spec}
\end{IEEEeqnarray}}
{\begin{multline}
S\left(f\right) = \sqrt{T} \times \\ \sum_{m=-\infty}^{\infty}\mathcal{J}_m^{1:K}\left(\{\alpha_k,-j\beta_k\}\right)\sinc\left[\pi T \left(f-\frac{m}{T}\right)\right].
\label{eq:MTSFM_Spec}
\end{multline}}
The AF of the MTSFM waveform, derived in Appendix \ref{sec:AppendixIII} , is expressed as
\ifthenelse {\boolean{singleColumn}}
 {\begin{multline}
\chi\left( \tau, \nu\right) = \left(\frac{T-|\tau|}{T}\right) \sum_{m, n}\mathcal{J}_m\left(\{\alpha_k,-j\beta_k\}\right)  \left(\mathcal{J}_n\left(\{\alpha_k,-j\beta_k\}\right) \right)^* \times \\ e^{-j\frac{\pi\left(m+n\right) \tau}{T}} \sinc\left[\pi\left(\dfrac{T-|\tau|}{T}\right)\left(\nu T + \left(m-n\right)\right)\right].
\label{eq:MTSFM_AF}
\end{multline}}
 {\begin{multline}
\chi\left( \tau, \nu\right) = \left(\frac{T-|\tau|}{T}\right) \times \\ \sum_{m, n}\mathcal{J}_m\left(\{\alpha_k,-j\beta_k\}\right)  \left(\mathcal{J}_n\left(\{\alpha_k,-j\beta_k\}\right) \right)^* \times \\ e^{-j\frac{\pi\left(m+n\right) \tau}{T}} \sinc\left[\pi\left(\dfrac{T-|\tau|}{T}\right)\left(\nu T + \left(m-n\right)\right)\right].
\label{eq:MTSFM_AF}
\end{multline}}
The ACF of the MTSFM is obtained by setting $\nu = 0$ and is expressed as 
\ifthenelse {\boolean{singleColumn}}
 {\begin{multline}
R\left(\tau\right) = \chi\left(\tau, \nu\right)|_{\nu=0} =\left(\dfrac{T-|\tau|}{T}\right)\sum_{m,n}\mathcal{J}_m^{1:K}\left(\{\alpha_k, -j\beta_k\}\right)\left(\mathcal{J}_n^{1:K}\left(\{\alpha_k,-j\beta_k\}\right)\right)^* \times \\e^{-j\frac{\pi\left(m+n\right)\tau}{T}} \sinc\left[\pi\left(\dfrac{T-|\tau|}{T}\right)\left(m-n\right)\right].
\label{eq:MTSFM_BAF_1}
\end{multline}}
{\begin{multline}
R\left(\tau\right) = \chi\left(\tau, \nu\right)|_{\nu=0} =\left(\dfrac{T-|\tau|}{T}\right) \times \\ \sum_{m,n}\mathcal{J}_m^{1:K}\left(\{\alpha_k, -j\beta_k\}\right)\left(\mathcal{J}_n^{1:K}\left(\{\alpha_k,-j\beta_k\}\right)\right)^* \times \\e^{-j\frac{\pi\left(m+n\right)\tau}{T}} \sinc\left[\pi\left(\dfrac{T-|\tau|}{T}\right)\left(m-n\right)\right].
\label{eq:MTSFM_BAF_1}
\end{multline}}
The result in \eqref{eq:MTSFM_AF} is a special case of that obtained by Auslander and Tolimieri \cite{Auslander}.  

The expressions \eqref{eq:MTSFM_Spec}-\eqref{eq:MTSFM_BAF_1} can be used to show that the MTSFM's waveform shape metrics possess contraction/expansion symmtery properties for varying pulse-length $T$ and swept bandwidth $\Delta f$ so long as the TBP = $T\Delta f$ remains fixed.  Consider a MTSFM waveform with TBP = $T \Delta f$ and modulation indices $\{\alpha_k, \beta_k\}$.  Now consider a second MTSFM waveform derived from the first with a new pulse-length $\tilde{T} = T/\xi$ and swept bandwidth $\tilde{\Delta f}=\xi \Delta f$ where $\xi$ is some non-zero scaling factor.  The scaling factor contracts or expand the waveform's duration and correspondingly will expand or contract the swept bandwidth in order to keep the TBP fixed  $\text{TBP}=\left(\frac{T}{\xi}\right) \xi \Delta f  = T\Delta f$.  The waveform's corresponding design coefficients $\{a_k, b_k\}$, which determine the waveform's swept bandwidth, are therefore scaled by $\xi$.  The second MTSFM waveform's resulting modulation indices are now expressed as 
\begin{align}
\tilde{\alpha}_k &= \frac{\xi a_k T}{k\xi} = \frac{a_k T}{k} = \alpha_k, \\
\tilde{\beta}_k &= \frac{\xi b_k T}{k\xi} = \frac{b_k T}{k} = \beta_k.
\end{align}
This means that for a fixed TBP and set of waveform modulation indices $\{\alpha_k, \beta_k\}$ the waveform shape characteristics of the MTSFM waveform possess the same structure but can be stretched or contracted in both duration and bandwidth.  Put another way, the modulation indices specify the waveform shape characteristics for a fixed TBP without explicitly defining the pulse-length $T$ or swept bandwidth $\Delta f$.  This property is loosely analogous to the way the order $N$ of a phase code is utilized to describe the waveform shape characteristics of PC waveforms with a specified TBP regardless of the pulse-length and bandwidth of the physical waveform that is transmitted \cite{Levanon, JianLiBookII}. 

The idea to utilize the MTSFM model explicitly for radar/sonar waveform synthesis \cite{HagueII, HagueIII} is new to the best of the author's knowledge.  However, the MTSFM waveform model itself is not entirely new and has been used for waveform analysis.  The MTSFM model appeared several times in the published literature dating back to the 1930's and 1940's when FM methods were being developed for analog communications systems.  Perhaps the most notable contribution to the published literature from that time is the work of Giacoletto \cite{Giacoletto} who used a similar model to \eqref{eq:MTSFM_1}-\eqref{eq:MTSFM_4} to analyze the spectrum of FM signals. There, the MTSFM waveform's spectrum was derived in closed-form using a product of sums of ordinary 1-D Bessel functions.  Work by \cite{subBandMod} utilized a CE-OFDM with waveform spectrum expressions similar to that of \cite{Giacoletto} to analyze both constant amplitude and spectral extent properties of simultaneously transmitted sonar waveforms.  Work by \cite{NLFMI} also used a MTSFM model in a form of paired echo analysis \cite{Cook} to analyze the the impact of Doppler effects on the ACF sidelobe structure of Non-Linear FM (NLFM) waveforms \cite{Levanon}.  Recent work by the author in \cite{HagueDiss, HagueI} used equations similar to \eqref{eq:MTSFM_1}-\eqref{eq:MTSFM_4} for the analysis of a family of thumbtack-like FM waveforms as well as several established waveforms in the literature.  There, exact closed form expressions were derived for the waveform's spectrum and AF using GBFs which to the best of the author's knowledge are all novel.   

The MTSFM belongs to a general class of multi-carrier waveforms.  The GBF-based representation of the MTSFM shown in \eqref{eq:MTSFM_5} and \eqref{eq:SFM_Fourier_Series} are a special case of the OFDM waveform model.  Additionally, the MTSFM representation given in \eqref{eq:MTSFM_1}-\eqref{eq:MTSFM_3} bears a particularly strong resemblance to CE-OFDM waveforms which has seen use in the radar community as a waveform for communication/radar spectrum sharing \cite{CEOFDM}.  CE-OFDM uses a standard OFDM signal as either the frequency or phase modulation function.  The data symbols serve as a set of constant modulus coefficients which are embedded into each orthogonal carrier in the modulation/phase function.  These carriers all share a common modulation index.  The data symbols can be utilized to not only to transmit data but to also as design coeffiicents that modify the resulting CE-OFDM's waveform shape properties.  The MTSFM's modulation and phase functions are also composed of orthogonal carriers except each carrier possesses its own variable modulation index.  However, some efforts in the literature have developed CE-OFDM models that bear a striking resemblance to the MTSFM model, particularly those of Sen and Nehorai \cite{Nehorai}, which analyzed the design of CE-OFDM waveforms for target detection in multi-path interference.  In fact, equations (12) and (13) from \cite{Nehorai} that describe their variant of CE-OFDM is essentially the MTSFM model with the addition of a common modulation index.  Further analysis of the results of \cite{Nehorai} showed that detection performance was not dependent upon either the number of coefficients or the values of those coefficients in the CE-OFDM model.  Thus, they concluded that their detector could not be improved by any efficient choice of the design coefficients.  As a result of this, all of their subsequent simulations proceeded to set all design coefficients to unity with a common modulation index thus resulting in the more traditional CE-OFDM waveform model. 
The primary contributions of this paper are that the MTSFM waveform model utilized in this paper uses equations \eqref{eq:MTSFM_1}-\eqref{eq:MTSFM_4} for waveform synthesis rather than analysis as was done in the previously mentioned efforts.  Additionally, this paper also provides novel exact closed-form expressions for the MTSFM waveform's AF and ACF using GBFs rather than a product of sums of 1-D Bessel functions which greatly simplifies analysis.  These equations may also be utilized in the analysis and synthesis of other multi-carrier waveform models, specifically the CE-OFDM waveform model.  

\subsection{Some Illustrative Design Examples}
\label{subsec:MTSFM_Ex}
\begin{figure}[ht]
\centering
\ifthenelse {\boolean{singleColumn}}
 {\includegraphics[width=1.0\textwidth]{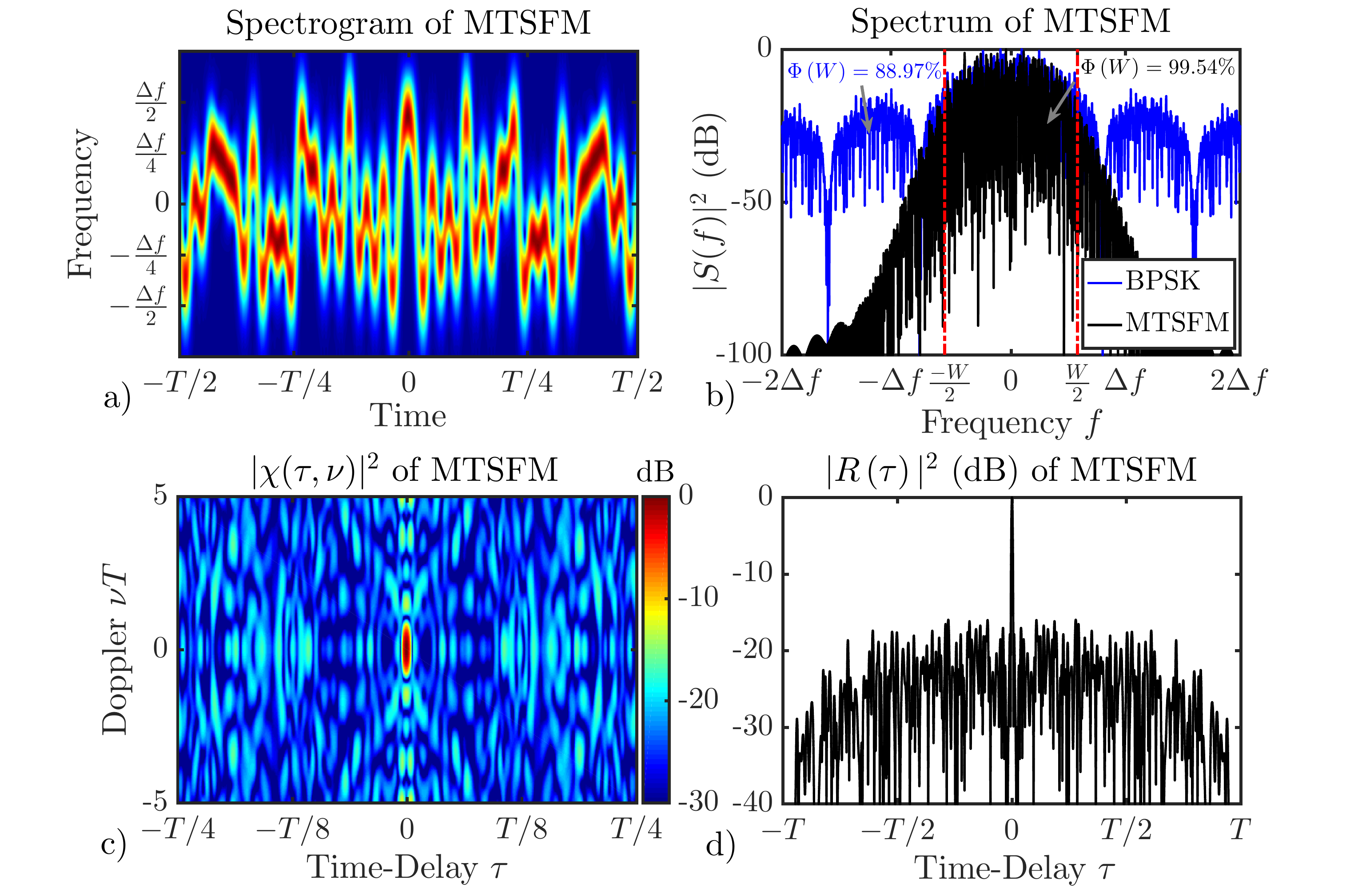}}
 {\includegraphics[width=0.5\textwidth]{MTSFM_Sonar_Journal_Paper.pdf}}
\caption{Spectrogram (a), spectrum (b), AF (c), and ACF (d) of an example MTSFM with TBP of 200.  The waveform is generated by initializing the Fourier design coefficients $a_k$ as i.i.d Gaussian random variables and scaled to occupy a desired swept bandwidth $\Delta f$.  The waveform resulting from this smooth pseudo-random modulation function possesses a thumbtack-like AF with a SE of $99.54\%$ across the band $W = \Delta f + 32/T$.  The spectrum of a PC with equivalent range resolution is also displayed in (b).  The PC waveform's spectrum has substantial spectral extent compared to the MTSFM resulting in a noticeably lower SE of $88.97 \%$ over the same band as that of the MTSFM.}
\label{fig:MTSFM_1}
\end{figure}

As mentioned earlier, while the MTSFM can synthesize a rich class of waveform types and AF shapes, this paper specifically focuses on the optimization and further refinement of thumbtack-like waveforms.  One efficient method to synthesize thumbtack-like MTSFM waveforms involves initializing the design coefficients $a_k$ and $b_k$ with i.i.d. Gaussian random variables as described in \cite{HagueIII}.  The resulting pseudo-random modulation function is continuous throughout its duration producing a spectrally compact thumbtack-like waveform.  Figure \ref{fig:MTSFM_1} shows the spectrogram, spectrum, AF, and ACF of an example MTSFM waveform whose modulation function is composed of $K = 32$ cosine harmonics resulting in an even-symmetric modulation function.  The corresponding waveform design coefficients $a_k$ are realized as i.i.d Gaussian random variables scaled so that the modulation function occupies a desired swept bandwidth $\Delta f$.  The resulting modulation indices $\alpha_k$ are shown in Table \ref{table:MTSFM_I} for reprodicibility purposes.  The waveform's TBP is 200.  For every waveform example in the paper, the waveform time-series is sampled at a rate $f_s = 10\Delta f$ and is tapered with a Tukey window with shape parameter $\alpha_T = 0.05$.  This mild tapering helps to notably reduce spectral leakage outside the waveform's swept bandwidth $\Delta f$ in exchange for a mild increase in PMEPR of 0.14 dB.  This is commonly employed in many sonar/ultrasound applications \cite{HagueIV, ultraSound_I} where the gradual ramping up of the waveform time-series amplitude helps to reduce distortion at the output of a piezoelectric transducer, a common artifact resulting from the transducer's transient response.  

\begin{table}[htb]
\caption[Modulation indices $\alpha_k$ used to generate the MTSFM shown in Figure \ref{fig:MTSFM_1}.]{Modulation indices $\alpha_k$ used to generate the MTSFM waveform shown in Figure \ref{fig:MTSFM_1}.}
\label{table:MTSFM_I}
\centering
\begin{tabular}{|c|c||c|c||c|c||c|c|}\hline
k   		&	$\alpha_k$	& k	  	& $\alpha_k$	& k   	 &	$\alpha_k$	& k	  	& $\alpha_k$\\\hline
1 		&	-4.2909		& 9 	&   1.2757		& 17	 &  -0.0817		& 25	&   0.3201      \\\hline
2      	&   2.5581			& 10	&   2.2940    	& 18	 &   1.3951		& 26	&  -0.0695     	\\\hline
3      	&  -2.4357    		& 11	&  -1.3832	   	& 19	 &   0.2267		& 27	&   0.0384      \\\hline 
4     	&  -2.7362    		& 12	&   0.0763    	& 20	 &  -0.1998		& 28	&  -0.6179      \\\hline
5   		&   4.8250	     		& 13	&  -0.0372      & 21	 &  -0.1366		& 29	&  -0.8159     	\\\hline
6   		&   0.3325	     		& 14	&   1.1292      & 22	 &   0.7981		& 30	&  -0.2587     	\\\hline
7   		&  -0.2497	     		& 15	&   0.7528      & 23	 &  -0.1766		& 31	&   0.4640      \\\hline
8   		&   1.5560	     		& 16	&   0.6234      & 24	 &   0.7064		& 32	&  -0.2167     	\\\hline
\end{tabular}
\end{table}  

From the figure, it is clear that the MTSFM's modulation function is smooth and without any transient-like artifacts in instantaneous frequency.  As a result of this, the majority of the waveform's energy is concentrated in its swept bandwidth $\Delta f$.  Using Carson's bandwidth rule, this MTSFM waveform should concentrate more than $98\%$ of its energy in a bandwidth $W = \Delta f + 32/T$.  Directly computing the waveform's SE using \eqref{eq:THETA} shows that this MTSFM waveform concentrates $99.54\%$ of its energy in that band.  The spectrum of a PC waveform employing a bi-phase code with the same TBP properties as the MTSFM is also show in Figure \ref{fig:MTSFM_1} and only achieves an SE value of $88.97\%$ with clearly visible spectral sidelobes which fall at a rate of 6 dB per octave.  The pseudo-random nature of the waveform's modulation function results in a waveform with a thumbtack-like AF.  This method of synthesizing families of thumbtack-like waveforms as was described in \cite{HagueIII} is generally robust and provided an efficient method to generate entire families of thumbtack-like MTSFM waveforms.  However, the MTSFM waveforms synthesized in \cite{HagueIII} were never optimized.  It is now the goal in this paper to modify the waveform design coefficients to further refine their performance characteristics.

The following design example demonstrates the ability of the MTSFM model to finely control waveform shape and assess the impact of increasing the number of design coefficients $K$.  In this scenario, the objective is to modify the waveform coefficients to reduce the sidelobe levels across a region of time-delays in the magnitude-square of the waveform's ACF $|R\left(\tau\right)|^2$.   The metric to be optimized is ACF's Integrated Sidelobe Ratio (ISR) with the constraint that the waveform's RMS bandwidth remains within $20\%$ of it's initial value.  Formally, the optimization problem can be stated as
\ifthenelse {\boolean{singleColumn}}
 {\begin{IEEEeqnarray}{rCl}
\underset{\alpha_{k}}{\text{min}}
\left[\dfrac{\int_{\Omega_{\tau}}|R\left(\tau\right)|^2  d\tau}{\int_{-\tau_m}^{\tau_m}|R\left(\tau\right)|^2  d\tau} \right]  \text{s.t.~} \left(1-\delta\right)\tilde{\beta}_{rms}^2 \leq \beta_{rms}^2\left(\{\alpha_k\}\right) \leq \left(1+\delta\right)\tilde{\beta}_{rms}^2
\label{eq:Problem1}
\end{IEEEeqnarray}}
{\begin{multline}
\underset{\alpha_{k}}{\text{min}}
\left[\dfrac{\int_{\Omega_{\tau}}|R\left(\tau\right)|^2  d\tau}{\int_{-\tau_m}^{\tau_m}|R\left(\tau\right)|^2  d\tau} \right] 
 \\ \text{s.t.~} \left(1-\delta\right)\tilde{\beta}_{rms}^2 \leq \beta_{rms}^2\left(\{\alpha_k\}\right) \leq \left(1+\delta\right)\tilde{\beta}_{rms}^2
\label{eq:Problem1}
\end{multline}}
where $0 \leq \delta < 1.0$ is a unitless parameter, $\tau_m$ denotes the first nulls of the ACF and therefore $2\tau_m$ is the ACF's null-to-null mainlobe width.  The $\tilde{\beta}_{rms}^2$ term is the initialized MTSFM waveform's RMS bandwidth.  The MTSFM's RMS bandwidth is expressed in terms of the modulation indices as \cite{HagueV}
\begin{IEEEeqnarray}{rCl}
\beta_{rms}^2 = \left(\frac{2\pi}{T}\right)^2 \sum_{k=1}^K k^2 \dfrac{\left(\alpha_k^2 + \beta_k^2\right)}{2}.
\label{eq:RMSBand}
\end{IEEEeqnarray}
For this example, the initial modulation indices are those shown in Table \ref{table:MTSFM_I}.  The region $\Omega_{\tau}$ where the ISR is to be optimized is  $\tau_m \leq |\tau| \leq 0.2T$ and $\delta = 0.2$.  This particular design problem is loosely analogous to adaptive beamforming where one wishes to reduce the sidelobes of the array response in a particular region while minimizing distortion elsewhere.  

The \emph{fmincon} function in MATLAB's Optimization Toolbox \cite{Matlab} is used to minimize \eqref{eq:Problem1} and all other waveform optimization methods described in this paper.  This optimization function utilizes a Sequential Quadratic Programming (SQP) method in order to handle the nonlinear constraints in \eqref{eq:Problem1}.  The routine does not guarantee convergence to a global minimum, but rather a local minimum.  It is also important to note that the implementation of this optimization routine is computationally heavy.  Optimizing either the modulus squared of the AF or ACF using the expressions in \eqref{eq:MTSFM_AF} and \eqref{eq:MTSFM_BAF_1} respectively are nonconvex quartic objective functions in the coefficients $c_m$ (i.e, the GBF's) which are notoriously computationally complex \cite{GladkovaI, GladkovaII}.  Clearly the optimization routine specified here are by no means effiicent and do not lend themselves to real-time operation.  However, recent efforts by \cite{Monga} have developed a quartic gradient descent algorithm that can be applied to shaping the AF while also being much less computationally expensive.  The task of developing efficient algorithms to optimize MTSFM waveforms is the topic of a future paper.  

This optimization problem was run four times each with a different number of design coefficients $K$.  The first run utilized the original $K=32$ coefficients for optimization. The subsequent three runs initialized the optimization problem with the 32 original design coefficients and then zero padded an additional 32, 64, and 96 coefficients resulting in $K = 64,~96,$ and $128$ coefficients respectively.  Increasing $K$ allows for more degrees of freedom in the problem and generally produces a waveform with better waveform shape characteristics.  However, there is a point of diminishing returns with increasing $K$.  Recall that the RMS bandwidth constraint represented by \eqref{eq:RMSBand} weighs higher order coefficients more heavily.  When running \eqref{eq:Problem1}, the RMS bandwidth tends to increase more rapidly with increasing $K$.  This results in the RMS bandwidth constraint being active during the optimization routine and limiting the values that $\alpha_k$ may take on.  Another consideration when increasing $K$ involves the waveform's SE.  The rate at which the spectral leakage of the waveform's spectrum falls off tends to have a small but noticeable impact on the waveform's SE.  This falloff rate decreases with increasing $K$.  Therefore, as $K$ is increased, the SE of the waveform can be reduced by a few percent.  

\begin{figure}[ht]
\centering
\ifthenelse {\boolean{singleColumn}}
 {\includegraphics[width=1.0\textwidth]{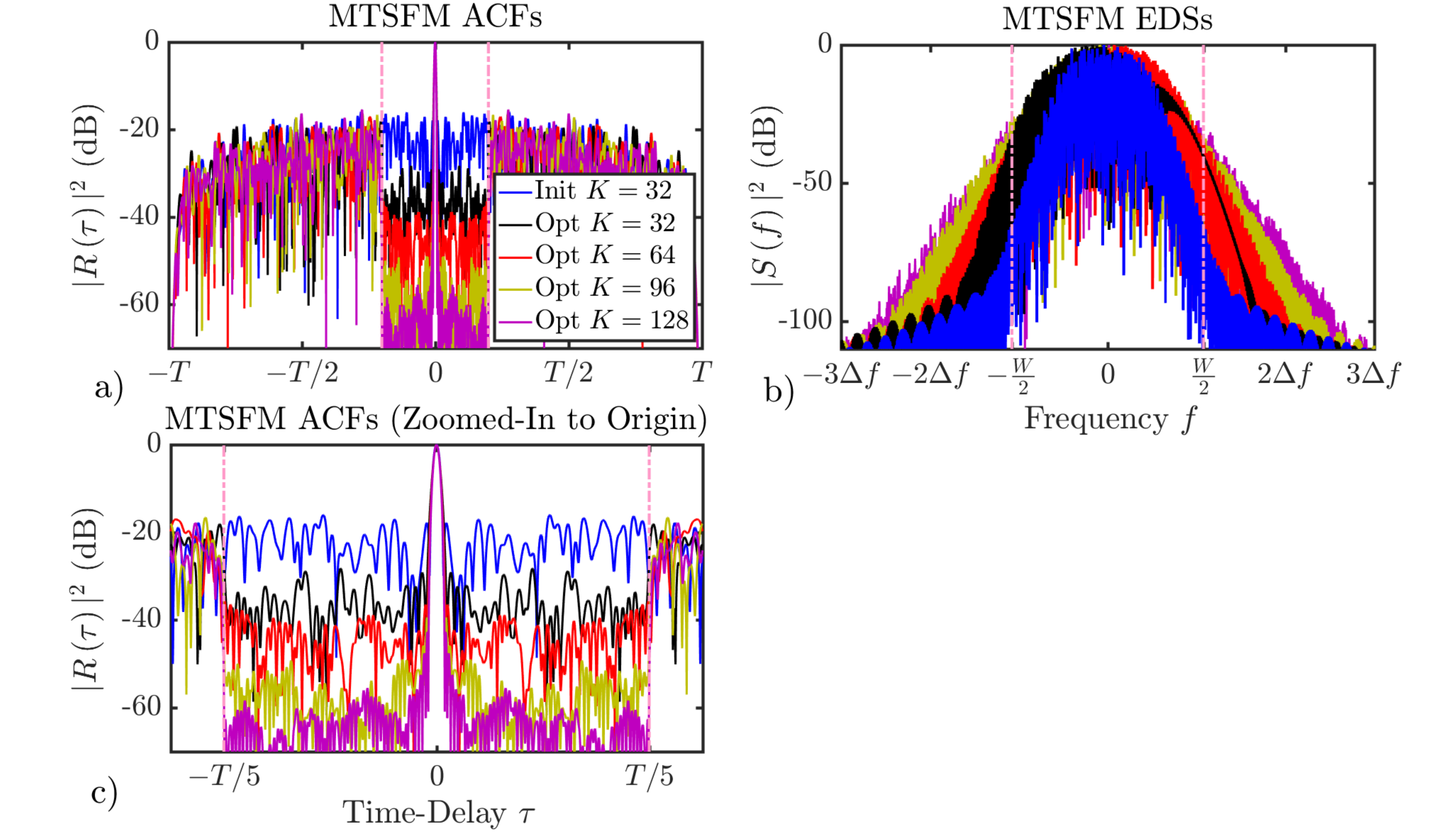}}
 {\includegraphics[width=0.5\textwidth]{illustrativeExample_Final.pdf}}
\caption{ACFs of the original and optimized waveforms displayed over their entire duration (a), their respective spectra (b), and their ACFs zoomed in at the origin (c). The optimized waveform ACFs possess drastically reduced sidelobes in the region $\Omega_{\tau}$ (denoted by the red dashed lines) without substantial distortion of the sidelobe and mainlobe structure outside the region $\Omega_{\tau}$.  Increasing $K$ resulted in further reduced ACF sidelobes at the expense of a reduced spectral leakage falloff rate resulting in lower SE compared to the initialized waveform.}
\label{fig:MTSFM_2}
\end{figure}

Figure \ref{fig:MTSFM_2} illustrates the results of this design problem.  As can be clearly seen in the figure, each optimal waveform with increased $K$ resulted in noticeably lower sidelobes over the region of time-delays $\Omega_{\tau}$.  Additionally, zooming in near the origin of the ACF shows that the mainlobe width of the optimal waveforms' ACFs have stayed essentially the same thus preserving the waveform's original range resolution.  This result is significant; usually the only option to reduce the sidelobe levels of waveforms with a thumbtack-like AF/ACF is to increase the waveform's TBP.  However, the waveforms shown in Figure \ref{fig:MTSFM_2} have their pulse-lengths fixed and the RMS bandwidth constraint ensured that the waveform did not sweep through a wider band of frequencies thus preserving the waveform's TBP.  Modifying the MTSFM's modulation indices $\alpha_k$ reduced the ACF pedestal over a region of time-delays without increasing the resulting waveforms' TBP.  

There is however a cost to increasing $K$.  As mentioned earlier, increasing $K$ tends to decrease the rate at which the waveform's spectral leakage falls off which results in a slightly reduced SE.  The decreased spectral falloff rate can be clearly seen in Figure \ref{fig:MTSFM_2} (b).  However, the reduction in SE is not severe.  Table \ref{table:MTSFM_II} lists several design characteristics of the resulting optimal MTSFM waveforms for each value of $K$ utilized in the optimization.  In addition to ISR and SE computed over $W = \Delta f + 32/T$, the reduction of ACF area $G$ is also displayed since the primary contributing factor to ISR improvement was the reduction of ACF sidelobe area component of the ISR metric.  The effect of increasing $K$ clearly had a substantial impact on improving $G$ and therefore ISR, especially for $K = 64$ and $96$.  The degree of improvement in ISR was less for $K=128$.  For the case where $K = 128$, it is likely that the RMS bandwidth constraint restricted the coefficients from being modified to further improve the ISR metric.  From Table \ref{table:MTSFM_II}, it is also clear that the optimal waveforms' SE was reduced slightly from the initial waveform which is due to the reduced falloff rate of the spectral leakage outside the swept bandwidth $\Delta f$.  

%\begin{table}[htb]
\begin{table}[htb]
\caption[List of performance metrics of the initial and Optimized MTSFMs.]{Reduction $G$ of ACF area over $\Omega_{\tau}$, ISR, and SE $\Theta\left(W\right)$ of the optimized MTSFMs using $K=32, 64, 96,$ and $128$ respectively.  As $K$ increases, the ISR is drastically improved.  However, the resulting waveforms' SE are less than the initial waveform's SE.}
\label{table:MTSFM_II}
\centering
\begin{tabular}{|c||c|c|c|}\hline
$K$   		&	$G$		& $\tilde{ISR}$ (dB) 	  & $\Theta\left(W\right) (\%)$       \\\hline
32 (Init) 	&	1.00		& -2.56				  & 99.54					\\\hline
32      	 	&  26.62		& -16.82     			  & 	96.81	    			  	  	\\\hline
64      		&  163.48       & -24.58       		  	  & 	95.59 	   				  	\\\hline 
96     		& 2148.12	     & -35.65	     			  & 	97.33        				\\\hline
128    		& 7870.37	     & -41.29				  & 97.39         				\\\hline
\end{tabular}
\end{table}  

The same principles can be applied to minimizing the MTSFM waveform's AF sidelobes over a region in range and Doppler.  As mentioned earlier, reducing the volume $V$ of a waveform's AF in a region $\Omega_{\tau, \nu}$ in the range-Doppler plane will accordingly reduce the sidelobe levels in that region.  The waveform design process should also implement constraints on the AF mainlobe structure such that it stays nearly the same width in both range and Doppler.  For MTSFM waveforms with an even-symmetric modulation function, the optimization problem can be stated as 
\ifthenelse {\boolean{singleColumn}}
 {\begin{equation}
\underset{\alpha_{k}}{\text{min}}
\left[\iint_{\Omega_{\tau, \nu}}|\chi\left(\tau, \nu\right)|^2  d\tau d\nu \right] 
\text{s.t.~} \left(1-\delta\right)\tilde{\beta}_{rms}^2 \leq \beta_{rms}^2\left(\{\alpha_k\}\right) \leq \left(1+\delta\right)\tilde{\beta}_{rms}^2
\label{eq:Problem2}
\end{equation}}
{\begin{multline}
\underset{\alpha_{k}}{\text{min}}
\left[\iint_{\Omega_{\tau, \nu}}|\chi\left(\tau, \nu\right)|^2  d\tau d\nu \right] \\
\text{s.t.~} \left(1-\delta\right)\tilde{\beta}_{rms}^2 \leq \beta_{rms}^2\left(\{\alpha_k\}\right) \leq \left(1+\delta\right)\tilde{\beta}_{rms}^2
\label{eq:Problem2}
\end{multline}}
where $\Omega_{\tau, \nu}$ is a sub-region of the range-Doppler plane excluding the mainlobe region.  The only AF mainlobe constraint is the RMS bandwidth.  This is because modifying the modulation indices $\alpha_k$ only influences the RMS bandwidth. The waveform's pulse-length stays fixed thus preserving the same mainlobe width in Doppler throughout the optimization routine.  

\begin{figure}[ht]
\centering
\ifthenelse {\boolean{singleColumn}}
 {\includegraphics[width=1.0\textwidth]{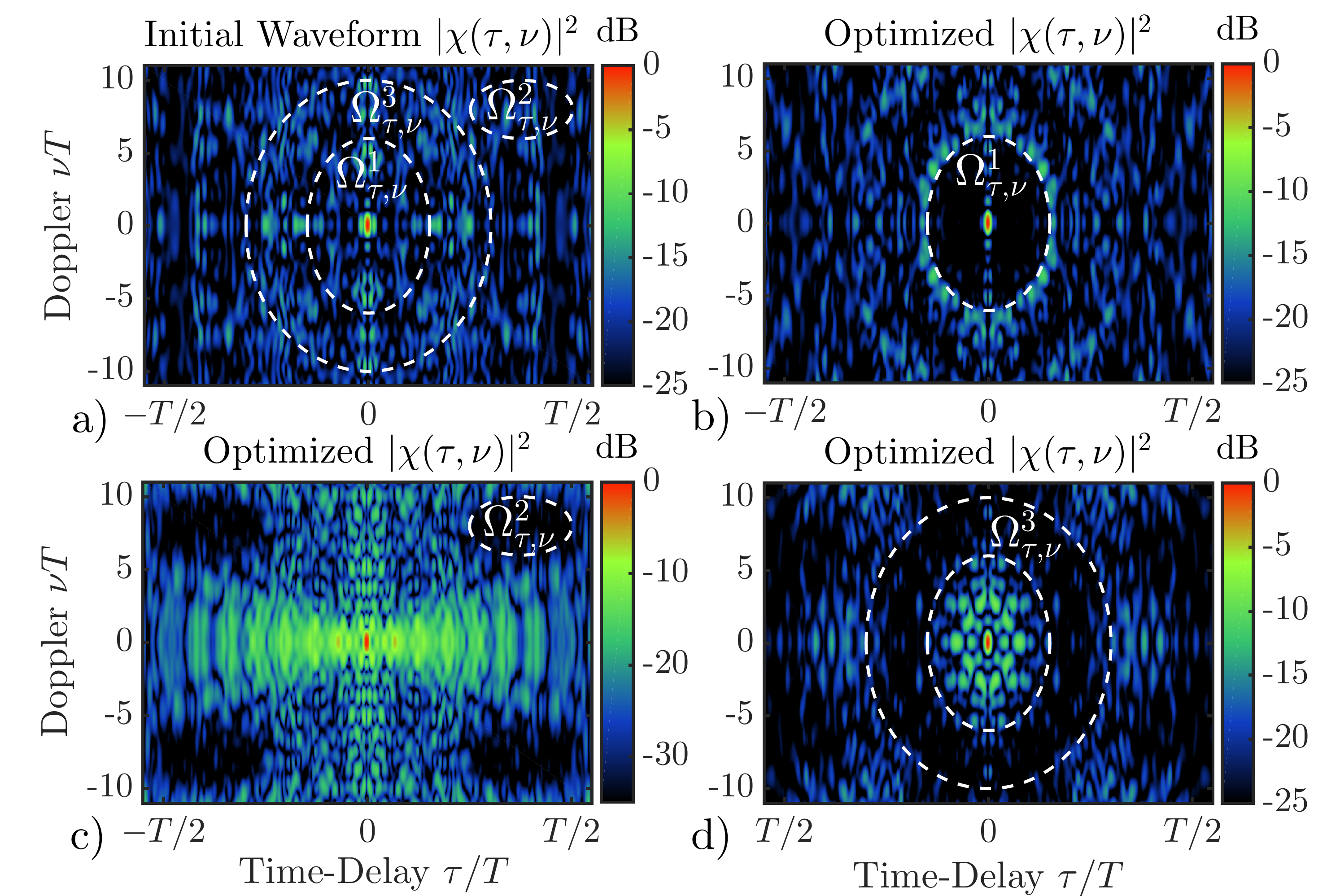}}
 {\includegraphics[width=0.5\textwidth]{volRegions.pdf}}
\caption{AF of the initialized MTSFM waveform (a) and resulting optimized waveforms' AFs (b)-(d) whose volume was minized over the three sub-regions in the range-Doppler plane.  While the resulting optimized AF regions are not completely volume free, the volume in each case was reduced by more than an order of magnitude.  This was achieved while keeping the mainlobe width of each AF essentially the same thus preserving the TBP of the initialized waveform.}
\label{fig:MTSFM_3}
\end{figure}

Figure \ref{fig:MTSFM_3} shows the initial MTSFM waveform and the result of running \eqref{eq:Problem2} on that initial waveform over three different ellipsoidally shaped regions $\Omega_{\tau, \nu}$ in the range-Doppler plane.  These ellipsoidally shaped regions were computationally shown to perform best with the thumbtack-like MTSFM waveform designs.  Each region is outlined by the white dashed lines in Figure \ref{fig:MTSFM_3}.  The first region, denoted $\Omega_{\tau, \nu}^1$ is an ellipse centered about the origin.  The second region denoted by $\Omega_{\tau, \nu}^2$ is an ellipse centered away from the origin.  The third region denoted $\Omega_{\tau, \nu}^2$ is an annulus centered about the origin.  Each of these regions were of area less than 4, which are necessary conditions for having volume free regions \cite{Hofstetter}.  In each case, while the optimized waveform's AF does not possess a completely volume free region, the volume in each of those regions were reduced by more than an order of magnitude.  This translated to reducing the sidelobe levels in those regions by more than 10 dB.  Most importantly, the mainlobe width in range and Doppler was not modified suggesting that the TBP has remained essentially fixed.  This shows that the MTSFM can be adapted to reduce the AF sidelobe pedestal over sub-regions in the range-Doppler place while retaining a fixed TBP product.  

\section{Performance Evaluation of the MTSFM Model}
\label{sec:MTSFM_Performance}
This section evaluates the MTSFM waveform model for the design and optimization of thumbtack-like waveforms and describes the metrics of performance for the waveform designs.  Specifically, this section describes the structure of the objective functions defined in \eqref{eq:Problem1} and \eqref{eq:Problem2} and compares optimized thumbtack-like MTSFM waveforms to thumbtack-like PC waveforms derived from design algorithms available in the published literature. 

\subsection{An Analysis of MTSFM Optimization Objective Functions}
\label{subsec:Methods}

\begin{figure}[ht]
\centering
\ifthenelse {\boolean{singleColumn}}
 {\includegraphics[width=1.0\textwidth]{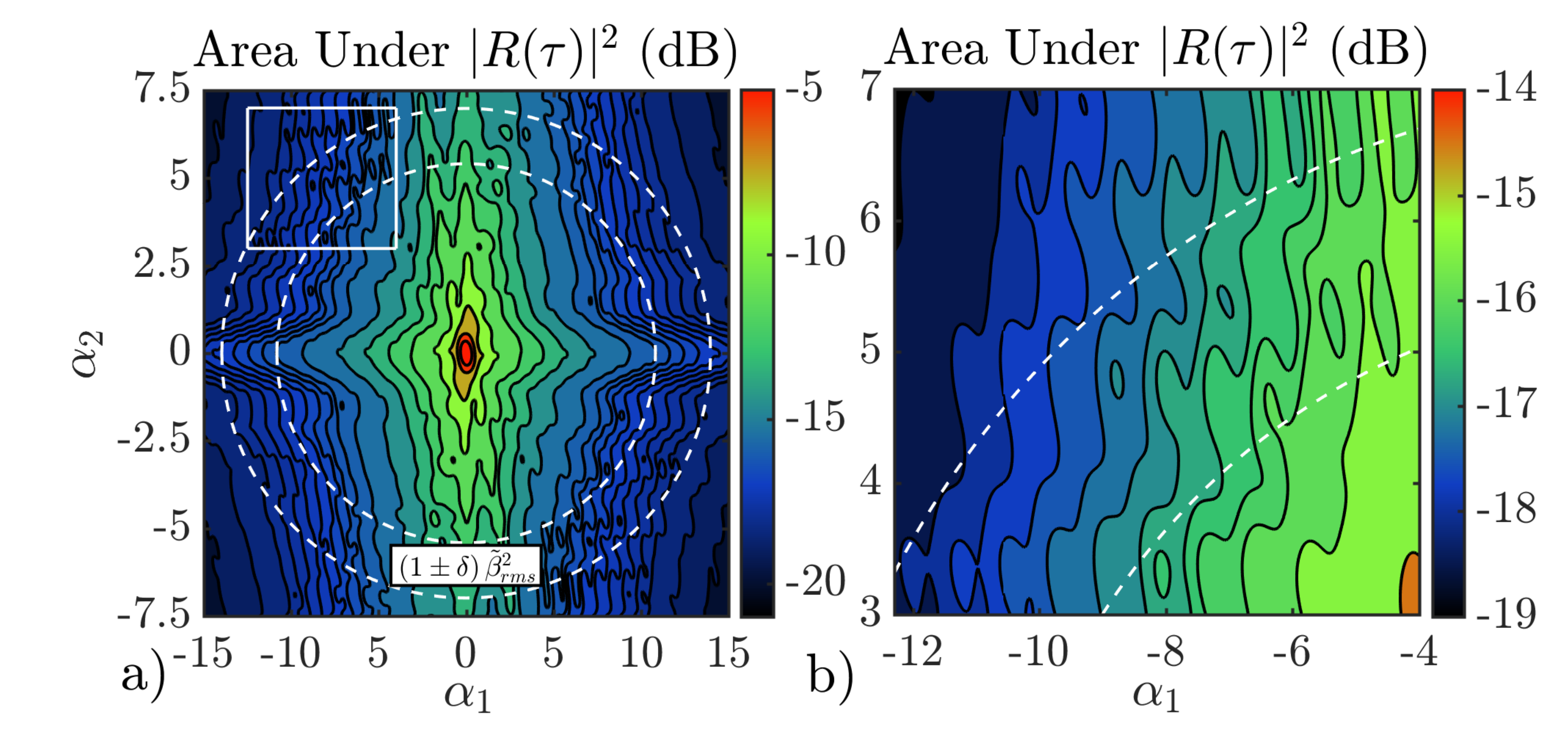}}
 {\includegraphics[width=0.5\textwidth]{basinOfAttraction.pdf}}
\caption{Area under $|R\left(\tau\right)|^2$ as a function of the two design parameters $\alpha_1$ and $\alpha_2$ (left panel) and a zoomed in version of the same plot (right panel) over the region depicted by the solid white box in (a).  This and many other MTSFM design objective functions are multi-modal and the initial values for $\alpha_1$ and $\alpha_2$ have a profound impact on the resulting optimal design.}
\label{fig:MTSFM_4}
\end{figure}

The design examples shown in Figures \ref{fig:MTSFM_2} and \ref{fig:MTSFM_3} show that the waveform design coefficients can be finely controlled to reduce ACF/AF sidelobes in a specified region of time-delays and Doppler values without compromising on mainlobe width.  However, each of the design examples are just one set of initial design coefficients.  These examples do not provide any insight into the structure of the objective functions described in \eqref{eq:Problem1} and \eqref{eq:Problem2}.  Figure \ref{fig:MTSFM_4} provides a simple visual of the structure of \eqref{eq:Problem1} and evaluates the area across all time-delays for a MTSFM waveform composed of two-tones with modulation indices $\alpha_1$ and $\alpha_2$ which are varied across a wide array of values. This produces a plot of ACF area as a function of $\alpha_1$ and $\alpha_2$. The plot in Figure \ref{fig:MTSFM_4} shows that there are multiple local extrema across a wide array of values for the modulation indices.  Depending on the initial values, the optimization routine will converge to different local minimums.  Similar results were obtained for the AF volume minimization problem defined in \eqref{eq:Problem2}.  This is likely due to the oscillatory nature of the GBFs.  Much like their 1-D counterparts, the GBFs of order $m$ and sums of GBFs over order $m$ have a highly oscillatory structure across the arguments $\{\alpha_k, \beta_k\}$ with specific regions of symmetry in the $K-$ dimensional plane \cite{Lorenzutta, Parker}.  This necessitates running a set of trials with waveforms whose initial modulation indices span across a wide variety of values to fully evaluate the effectiveness of the resulting waveform designs derived from either of the MTSFM optimization problems defined in \eqref{eq:Problem1} and \eqref{eq:Problem2}.

The following simulation generated 100 realizations of MTSFM waveforms with both even and odd symmetry in their modulation functions.  The waveforms possess a TBP of 200 and are composed of 32 modulation indices initialized using the method developed in reference \cite{HagueII}.  The optimization problem described in \eqref{eq:Problem1} was run this time to optimize the ISR metric across all time-delays.  Like the ISR optimization examples from the last section, the sidelobe region area had the most profound impact on minimizing the ISR as the RMS bandwidth barely varied for any of the trials.  Therefore, an effective measure of performance of these simulations is to directly analyze ACF area.  The analysis of these trials use two performance metrics. The first is the reduction of area of each trial denoted as $G_i$ and is expressed as
\begin{equation}
G_i = \dfrac{A_0\left(i\right)}{A_{opt}\left(i\right)}
\label{eq:G}
\end{equation}
where $A_0\left(i\right)$ and $A_{opt}\left(i\right)$ are the area of the initial and optimized waveforms.  Since the initial waveforms' modulation indices are randomly initialized, the initial areas over $\Omega_{\tau}$ are different for each waveform trial.  The metric $G_i$ therefore only gives a partial description of performance improvement. To account for the variation in initial area for each waveform trial, these simulations also measure a normalized version of area reduction denoted as $\tilde{G}_i$ and is expressed as
\begin{equation}
\tilde{G}_i = \dfrac{A_0\left(i\right)/A_{opt}\left(i\right)}{A_0\left(i\right)/\min{\{A_{opt}\}}} = \dfrac{\min\{A_{opt}\}}{A_{opt}\left(i\right)}
\label{eq:Norm}
\end{equation}
where $\min\{A_{opt}\}$ is the lowest area of all the 100 optimized waveforms for that set of waveform trials. Ideally $\min\{A_{opt}\}$ should be the global minimum of \eqref{eq:Problem1} that satisfies the RMS bandwidth constraints. However, since this value is unknown, the minimum from the 100 trial waveforms is used instead. 

\begin{figure}[ht]
\centering
\ifthenelse {\boolean{singleColumn}}
 {\includegraphics[width=1.0\textwidth]{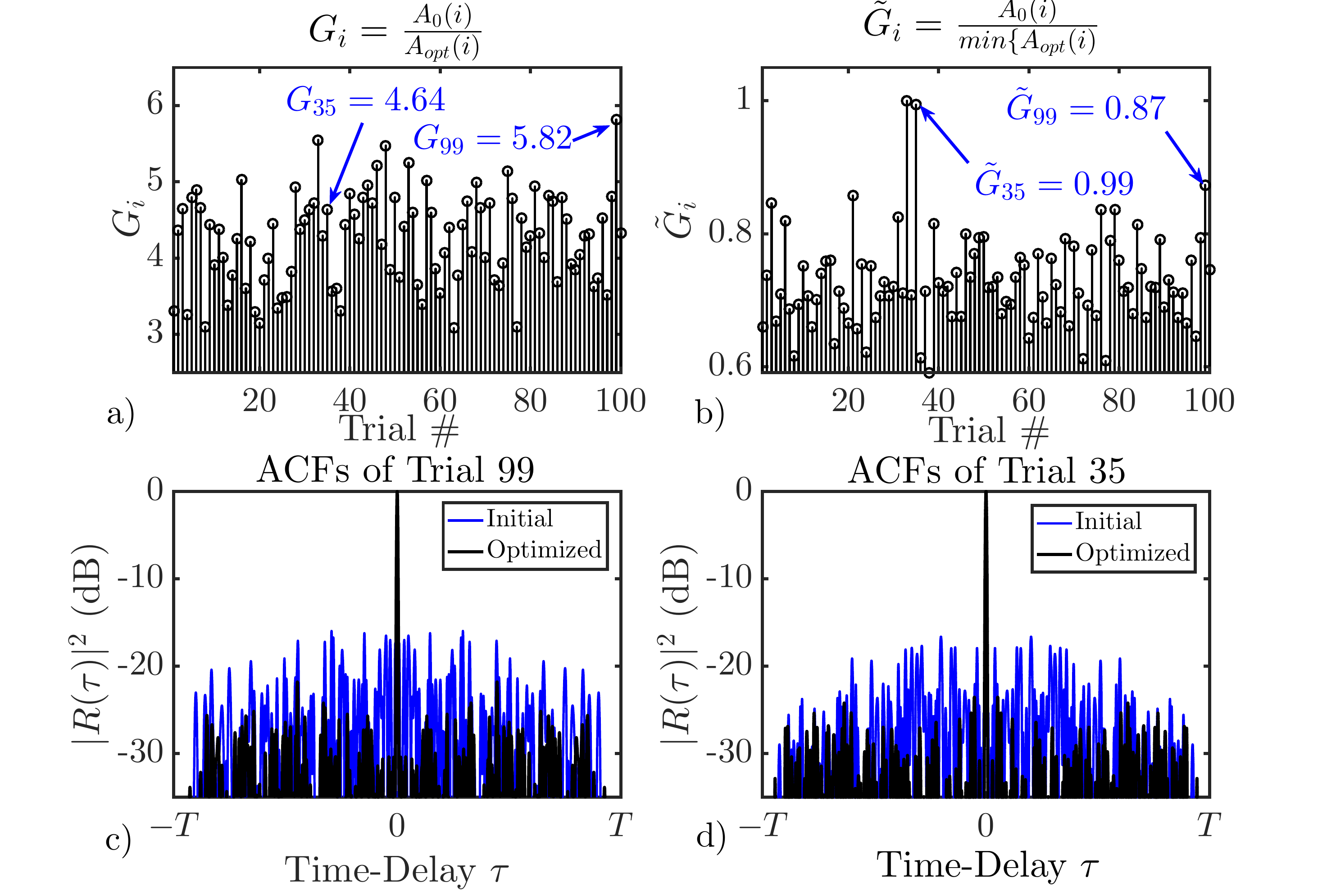}}
 {\includegraphics[width=0.5\textwidth]{normalizedGainExample.pdf}}
\caption{Area reduction $G_i$ (a) and normalized area reduction $\tilde{G}_i$ (b) for the 100 waveform trials. The initial and optimized ACFs of waveforms 99 and 35 are displayed in (c) and (d) respectively. Waveform 99 had the greatest reduction in area, but started with a larger initial area than waveform 35. On the other hand, waveform 35 was initialized with lower ACF area than waveform 99 and its resulting optimized version achieved the lowest ACF area overall.}
\label{fig:MTSFM_5}
\end{figure}

Figure \ref{fig:MTSFM_5} shows the area reduction $G_i$ and normalized area reduction $\tilde{G}_i$ for the even-symmetric MTSFM trials. Additionally, the initial and optimized waveform ACFs from two of the trials (waveforms 35 and 99 respectively) are also displayed. The optimal designs possess ACF areas that are on average 4.24 times lower than their initialized versions. The greatest area reduction was 5.82 achieved by waveform 99 resulting in an ISR of -7.47 dB.  However, waveform 99 only achieved a $\tilde{G}_i$ of 0.87 implying that its ACF area was not the lowest of all the trials.  Waveform 35 on the other hand, which achieved an area reduction of only 4.64, achieved a lower ISR of -8.32 dB $\tilde{G}_i$ of 0.99, much closer to the lowest ACF area value of the trial waveforms.  This means that waveform 35 achieved a lower overall ACF area than waveform 99 even though waveform 99 achieved the greatest reduction in ACF area $G_i$.  This is because waveform 35 was initialized with coefficients that were close to a local minimum in the ACF area objective function that was lower than the region where waveform 99 was initialized.  This can even be seen in Figure \ref{fig:MTSFM_5} where the ACF sidelobes of waveform 99 are on average slightly higher than waveform 35's, specifically for time-delays greater than $|\tau| \geq 0.5T$.  This single set of trials demonstrates the multi-modal structure of the multi-dimensional ACF area objective function.  

\begin{figure}[ht]
\centering
\ifthenelse {\boolean{singleColumn}}
 {\includegraphics[width=1.0\textwidth]{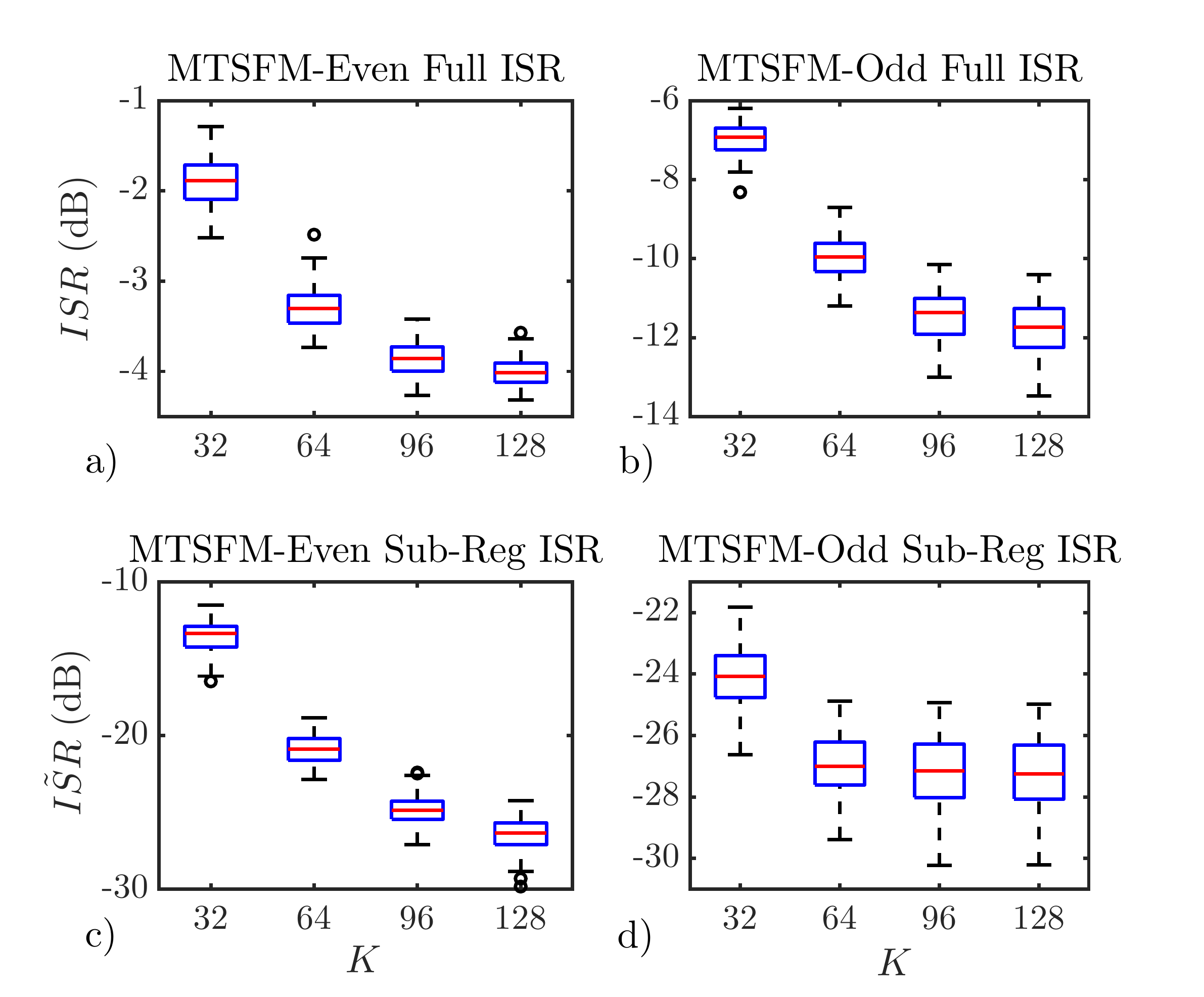}}
 {\includegraphics[width=0.5\textwidth]{varying_K_Trials_Box_Plot.pdf}}
\caption{Box plots of ISR (a) and $\tilde{ISR}$ (b) versus $K$ for MTSFM waveforms with even and odd symeetric modulation functions respectively.  Optimized MTSFM waveforms with odd symmetry tend to have substantially lower ACF sidelobes than MTSFMs with even-symmetry.}
\label{fig:MTSFM_6}
\end{figure}

Figure \ref{fig:MTSFM_6} shows box plots of resulting ISR values for a set of 100 trials of MTSFM waveforms with even and odd symmetry in their modulation functions and an increasing number of design coefficients $K$.  Both the ISR for all time-delays (denoted as Full ISR) and a sub-region $\tau_m \leq \Omega_{\tau} \leq 0.2T$ of time-delays (denoted as Sub-Reg ISR) were computed for each MTSFM waveform type.  The box in each box plot represents the 2nd and 3rd quartile of the trial data.  The whiskers represent the inner fence of the data (i.e 1.5 times the inter-quartile range (IQR)).  The circles denote statistical outliers in the results for each trial.  The waveforms were optimized using \eqref{eq:Problem1}.  As was explained in the previous section, increasing $K$ increases the degrees of freedom that the objective function may explore which generally results in more refined waveform designs at the expense of a slightly reduced SE.  The results demonstrate the clear advantage of increasing $K$ up to a point of diminishing returns.  As mentioned earlier, for \eqref{eq:Problem1}, these diminishing returns are a result of the RMS bandwidth constraint restricting the design coefficients from achieving further optimal designs.  The results also show that the odd-symmetric MTSFM waveforms generally possess notably lower ACF sidelobes.  The even-symmetric MTSFM only shows nearly comparable performance for large $K$ for the sub-region ISR metric.  The intuition for why this occurs can be derived from considering the structure of the modulation functions for the two versions of MTSFM.  The odd-symmetric modulation functions, while still possessing a thumbtack-like AF, have small but non-zero coupling between their range and Doppler measurements \cite{Cook, Ricker}.  This coupling has the effect of shearing the AF volume out to non-zero Doppler values in a manner loosely analogous to how a LFM waveform shears the AF volume of a simple pulse out to high non-zero Doppler values \cite{Levanon}.  It is likely that odd-symmetric MTSFM designs can exploit this characteristic to reduce ACF sidelobes more aggressively than the even-symmetric MTSFM.  This highlights the primary difference between even and odd symmetric MTSFM modulation functions, the odd-symmetric MTSFM has the ability to shear AF volume to non-zero Doppler values.  

\subsection{Comparing the MTSFM to other Waveform Optimization Methods}
\label{subsec:ACF_ISR}
The previous sections demonstrated the ability of the MTSFM model to adapt its waveform shape characteristics by modifying the modulation indices $\{\alpha_k, \beta_k\}$ and described the structure and behavior of the objective functions derived from \eqref{eq:Problem1} and \eqref{eq:Problem2}.  While the MTSFM has a clear advantage in higher SE compared to standard PC waveforms, it is not clear how optimized thumbtack-like MTSFM waveforms compare to PC waveforms designed for the same application.  This section explores this comparison by running a set of optimization trials of thumbtack-like MTSFM and PC waveforms across four TBP values of 32, 64, 128, and 256.  The TBP values were chosen since the number of chips $N$ defines the TBP and the algorithms used to generate the phase-codes used in this analysis require a value of $N$ that is a power of two.   The waveform design trials analyze even and odd MTSFM waveforms and used initial modulation indices that generated thumbtack-like waveforms.  Two forms of PC waveforms were used to compare to the two variants of MTSFM waveforms.  The time-series model for a PC waveform is expressed as 
\begin{IEEEeqnarray}{rCl}
s_{pc}\left(t\right) = \sum_{i=1}^Na\left(t-iT/N\right) e^{j2\pi f_c t + \theta_i}
\end{IEEEeqnarray}
where $a\left(t-iT/N\right)$ is the real-valued and positive amplitude tapering function of each chip in the PC waveform and $\theta_i$ is the phase of each chip (i.e, the phase code) of the PC waveform.  

PC waveforms using Maximal-Length Shift Register (MLSR) sequences, also known simply as M-Sequences, where used to compare against even MTSFM waveforms.  PC waveforms designed using the Cyclic-Algorithm New (CAN) algorithm \cite{jianLiIII, JianLiBookII, jianLiI} which also leverages an ISR-like metric for optimizing phase-codes, were used to compare against odd-symmetric MTSFM waveforms.  It is important to note that the ISR figure of merit defined by \cite{jianLiIII, JianLiBookII} only computes the sum of squares of the phase-code sidelobes at discrete points in time, rather than directly compute the ISR as defined in \eqref{eq:Problem1}.  This paper evaluates the ISR as defined in \eqref{eq:Problem1} of the physical PC waveform in order to provide a fair comparison between the optimized MTSFM and PC waveforms.  A similar analysis comparison can be performed for the AF volume metric over sub-regions in the range-Doppler plane.  However, as mentioned earlier, there exist strict bounds on the size of clear regions of the AF.  Both waveforms exhibit essentially the same ability to suppress AF volume; as long as the region $\Omega_{\tau, \nu}$ is of area less than 4, the AF's possessed essentially no volume in the region except for the volume contribution from the mainlobe.  This bound is not waveform specific and therefore does not provide a meaningful comparison between the MTSFM and PC waveforms.  However, since the ISR, and more specifically ACF area, do not follow such strict bounds, it is more likely that ISR will provide a more meaningful comparison between the two waveform types.

\begin{figure}[ht]
\centering
\ifthenelse {\boolean{singleColumn}}
 {\includegraphics[width=1.0\textwidth]{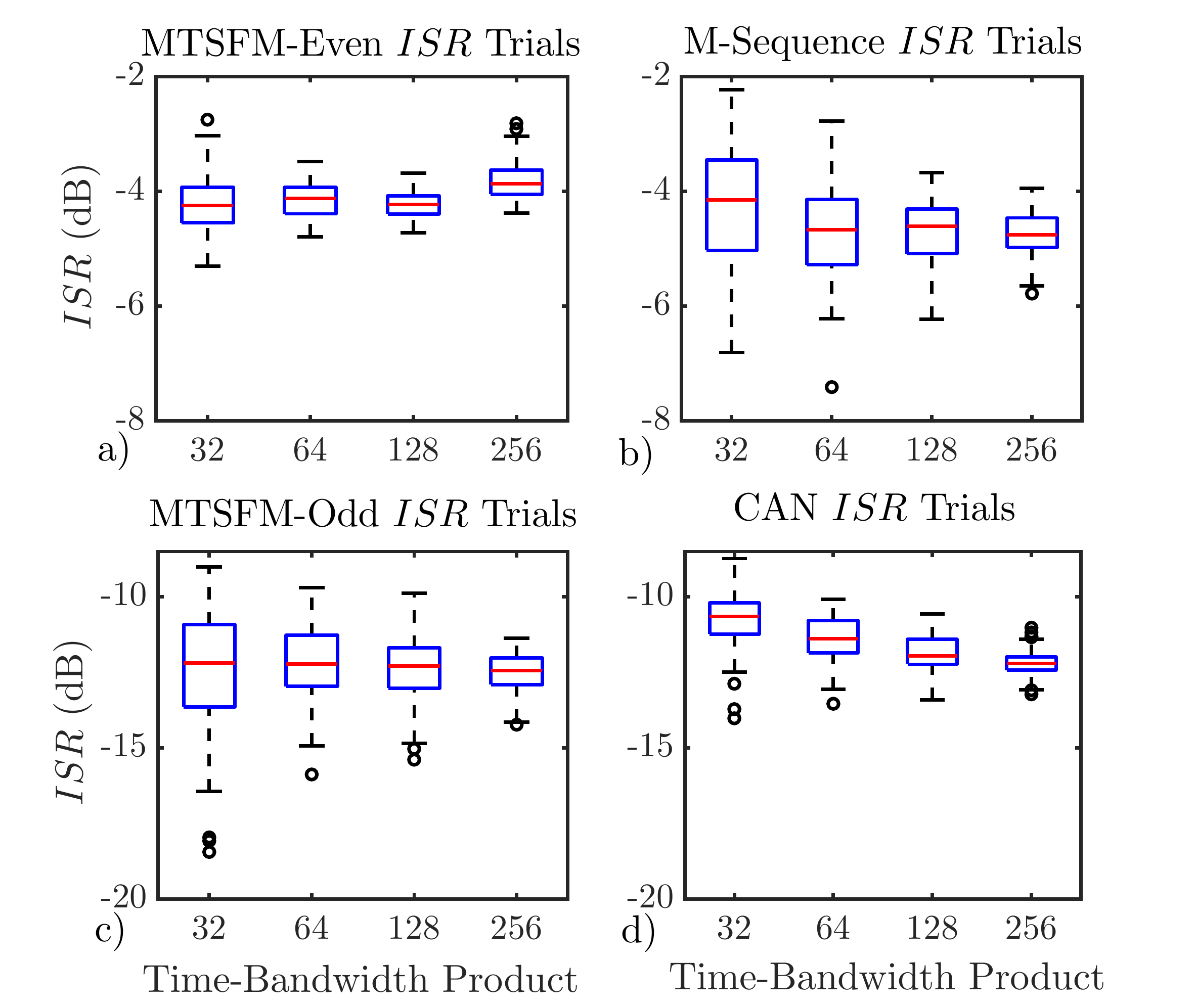}}
 {\includegraphics[width=0.5\textwidth]{thumbtackTrialBoxPlots.pdf}}
\caption{Box plot of ISR values for the Even/Odd symmetric MTSFM waveform trials compared against M-Sequence and CAN optimized PC waveforms across four different TBP values.  The even-symmetric MTSFM performs similarly to that of a PC waveform with an M-Sequence code across TBPs.  The odd-symmetric MTSFM waveforms on average out performed the CAN optimized PC waveforms for smaller TBPs.}
\label{fig:MTSFM_7}
\end{figure}

Figure \ref{fig:MTSFM_7} shows box plots of ISR values derived from 100 trials for each of the four waveform types and four TBP values.  The even MTSFM and M-Sequence based PC waveforms performed very closely across TBP, though the M-Sequence based PC waveforms display larger variation in ISR and the even MTSFM median ISR values were slightly higher.  The opposite behavior is observed for the odd MTSFM and CAN based PC waveforms.  For all TBPs, the odd MTSFM's median ISR was less than or equal to that of the CAN based PC waveforms.  However, the odd MTSFMs also display greater variation in ISR across all TBPs.  Overall, the MTSFM's ISR is at least competitive with and at times better than the PC waveform design methods.  This coupled with their spectral efficiency and constant envelope makes the MTSFM a potentially attractive adaptive waveform design model. 

\section{Conclusion}
\label{sec:Conclusion}
This paper introduced the MTSFM model as an adaptive FM waveform design method that synthesizes constant envelope and spectrally compact waveforms that are well suited for transmission on practical transmitter electronics. The MTSFM waveforms' modulation function is represented as a finite Fourier series expansion where the Fourier coefficients are utilized as a finite discrete set of design parameters.  These design parameters are adjusted to modify the waveform shape characteristics of the waveform.  The MTSFM has an exact mathematical definition for its time-series using GBFs which allow for deriving analytical expressions for the the MTSFM's waveform shape characteristics.  These expressions allow for establishing well-defined optimization problems that finely tune the MTSFM's properties while naturally possessing the constant envelope and high SE properties necessary for efficient transmission on realistic transmitter electronics.

The primary goal of this paper was to demonstrate the fundamental properties of the MTSFM waveform model and demonstrate them via illustrative design examples.  Simulations specifically focused on the design of thumbtack-like waveforms and demonstrated the MTSFM’s ability to reduce area or volume in a specified region of the waveform’s AF or ACF respectively.   This is accomplished while minimizing the distortion elsewhere in the AF/ACF and maintaining the initialized waveform's TBP.  The performance characteristics of the MTSFM are competitive with other optimal PC waveform design methods in terms of their ACF shapes while clearly out-performing PC waveforms with a noticeably higher SE.  

Thumbtack-like waveforms were chosen for this analysis as they are perhaps the simplest waveform type to demonstrate many of the properties that an adaptive waveform model like the MTSFM possesses.   However, there are likely numerous other problems of interest to the radar and sonar communities where the MTSFM model may be applicable.  The GBF-based representation of the MTSFM waveform in \eqref{eq:MTSFM_5} and \eqref{eq:SFM_Fourier_Series} establishes a mathematically precise and convenient way to describe the MTSFM waveform shape properties.  Due to the MTSFM's strong resemblance to other multi-carrier waveform models such as OFDM and CE-OFDM, the GBF-based representation may well provide insight into analysis and adaptive synthesis of these multi-carrier waveforms.  The waveform design methods described in this paper can be readily extended to design waveforms that possess non-zero range-Doppler coupling, also known as Doppler tolerant waveforms.  Optimizing these waveform types result in NLFM waveforms with finely tuned ACF properties with very low sidelobe leves.  This was demonstrated in \cite{HagueVI} and will be discussed in greater detail in an upcoming paper.   Lastly, this paper focused on optimizing a single waveform's design characteristics.  Many systems employ entire families of waveforms with specific ACF and Cross-Correlation Function (CCF) properties with one another.  Such a problem requires the optimization of a multi-objective function with a greater number of design parameters.  The MTSFM was recently applied to this problem in \cite{HagueVII} and will be investigated in greater detail in another upcoming paper.  

% or


% or
\appendices

\section{The MTSFM and the GBF Jacobi-Anger Expansion}
\label{sec:AppendixI}
Starting with the complex Fourier series representation in \eqref{eq:MTSFM_5} and making the substitution $\theta = \frac{2\pi t}{T}$ where $-\pi \leq \theta \leq \pi$ results in the expression 
\begin{IEEEeqnarray}{rCl}
s\left(\theta\right) = \dfrac{\rect\left(\theta/2\pi\right)}{\sqrt{2\pi}}\sum_m c_m e^{j\theta},
\label{eq:App_A_1}
\end{IEEEeqnarray}
resulting in a general complex Fourier series with period $2\pi$.  Solving for the complex Fourier series coefficients $c_m$ results in the integral expression
\ifthenelse {\boolean{singleColumn}}
{\begin{IEEEeqnarray}{rCl}
c_m = \dfrac{1}{2\pi}\int_{-\pi}^{\pi}\exp\Biggl\{j\left[m\theta - \sum_{k=1}^K \alpha_k \sin\left(k \theta\right) -\beta_k\cos\left(k\theta\right) \right]\Biggr\}d\theta,
\label{eq:App_I_2}
\end{IEEEeqnarray}}
{\begin{multline}
c_m = \dfrac{1}{2\pi}\int_{-\pi}^{\pi}\exp\Biggl\{j\left[m\theta - \sum_{k=1}^K \alpha_k \sin\left(k \theta\right) \right. \\ \left. -\beta_k\cos\left(k\theta\right) \right]\Biggr\}d\theta,
\label{eq:App_I_2}
\end{multline}}
which is the integral representation of the MT-GBF \cite{Dattoli, DattoliII}.  Thus, using \eqref{eq:App_I_2}  and re-substituting $\theta = \frac{2\pi t}{T}$ back into \eqref{eq:App_A_1}, the complex Fourier series representation for the MTSFM model in \eqref{eq:MTSFM_4} is expressed as 
\begin{IEEEeqnarray}{rCl}
s\left(t\right) = \dfrac{\rect\left(t/T\right)}{\sqrt{T}}\sum_{m=-\infty}^{\infty} \mathcal{J}_m^{1:K}\left(\{\alpha_k; -j\beta_k\}\right)e^{\frac{j2\pi m t}{T}}.
\label{eq:App_I_3}
\end{IEEEeqnarray}
The result in \eqref{eq:App_I_3} may also be derived via inspection of the MT-GBFs generating function \cite{DattoliII}
\ifthenelse {\boolean{singleColumn}}
{\begin{IEEEeqnarray}{rCl}
\exp\Biggl\{\frac{1}{2}\sum_{k=1}^Kx_k\left(\ell^k - \frac{1}{\ell^k}\right)+y_k\left(\ell^k + \frac{1}{\ell^k}\right)\Biggr\} = \sum_{m=-\infty}^{\infty}\mathcal{J}_m^{1:K}\left(\{x_k; y_k\}\right)\ell^m.
\label{eq:App_I_4}
\end{IEEEeqnarray}}
{\begin{multline}
\exp\Biggl\{\frac{1}{2}\sum_{k=1}^Kx_k\left(\ell^k - \frac{1}{\ell^k}\right)+y_k\left(\ell^k + \frac{1}{\ell^k}\right)\Biggr\} \\ = \sum_{m=-\infty}^{\infty}\mathcal{J}_m^{1:K}\left(\{x_k; y_k\}\right)\ell^m.
\label{eq:App_I_4}
\end{multline}}
Setting $x_k = \alpha_k$, $y_k = -j\beta_k$, and $\ell = e^{j\theta}$ yields the Jacobi-Anger identity for MT-GBFs
\ifthenelse {\boolean{singleColumn}}
{\begin{IEEEeqnarray}{rCl}
\exp\Biggl\{j\sum_{k=1}^K\alpha_k\sin\left(k\theta\right)-\beta_k\cos\left(k\theta\right)\Biggr\} = \sum_{m=-\infty}^{\infty}\mathcal{J}_m^{1:K}\left(\{\alpha_k; -j\beta_k\}\right)e^{jm\theta}.
\label{eq:App_I_5}
\end{IEEEeqnarray}}
{\begin{multline}
\exp\Biggl\{j\sum_{k=1}^K\alpha_k\sin\left(k\theta\right)-\beta_k\cos\left(k\theta\right)\Biggr\} = \\ \sum_{m=-\infty}^{\infty}\mathcal{J}_m^{1:K}\left(\{\alpha_k; -j\beta_k\}\right)e^{jm\theta}.
\label{eq:App_I_5}
\end{multline}}
Finally, setting $\theta = \frac{2\pi}{T}$ results in the complex Fourier series representation of the MTSFM waveform model.
\begin{align}
s\left(t\right) &= \exp\Biggl\{j\sum_{k=1}^K\alpha_k\sin\left(\frac{2\pi k t}{T}\right)-\beta_k\cos\left(\frac{2\pi k t}{T}\right)\Biggr\}\IEEEnonumber \\   &= \sum_{m=-\infty}^{\infty}\mathcal{J}_m^{1:K}\left(\{\alpha_k; -j\beta_k\}\right)e^{j\frac{2\pi m t}{T}}.
\label{eq:App_I_6}
\end{align}

For the case of a MTSFM waveform with an even-symmetric modulation function, the odd modulation indices $\beta_k$ are all zero.  The representation in \eqref{eq:App_I_6} still holds, but now the complex Fourier series coefficients are cylindrical GBFs with arguments $\{\alpha_k\}$.  The cylindrical GBF has a similar integral representation as \eqref{eq:App_I_2} as well as generating function and Jacobi-Anger identity but with only $\{\alpha_k\}$ as arguments.  For the case of a MTSFM waveform with an odd-symmetric modulation function, the even modulation indices $\alpha_k$ are all zero.  The representation in \eqref{eq:App_I_6} then uses $K$-dimensional M-GBFs.  This type of GBF again has a similar integral expression as the other two versions but has a modified generating function
\begin{IEEEeqnarray}{rCl}
\exp\Biggl\{\frac{1}{2}\sum_{k=1}^K\beta_k\left(\ell^k + \frac{1}{\ell^k}\right)\Biggr\} = \sum_{m=-\infty}^{\infty}\mathcal{I}_m^{1:K}\left(\{\beta_k\}\right)\ell^m.
\label{eq:App_I_7}
\end{IEEEeqnarray}
Letting $\ell = e^{j\theta}$ and setting $z_k = -j\beta_k$ yields the Jacobi-Anger identity for $K$ dimensional M-GBFs
\ifthenelse {\boolean{singleColumn}}
{\begin{IEEEeqnarray}{rCl}
\exp\Biggl\{-j\sum_{k=1}^K\beta_k\cos\left(\frac{2\pi k t}{T}\right)\Biggr\} = \sum_{m=-\infty}^{\infty}\mathcal{I}_m^{1:K}\left(\{-j\beta_k\}\right)e^{j\frac{2\pi m t}{T}}.
\label{eq:App_I_8}
\end{IEEEeqnarray}}
{\begin{multline}
\exp\Biggl\{-j\sum_{k=1}^K\beta_k\cos\left(\frac{2\pi k t}{T}\right)\Biggr\} \\ = \sum_{m=-\infty}^{\infty}\mathcal{I}_m^{1:K}\left(\{-j\beta_k\}\right)e^{j\frac{2\pi m t}{T}}.
\label{eq:App_I_8}
\end{multline}}

Thus, the MTSFM's complex Fourier coefficients can be expressed in exact closed form in terms of different version of GBFs depending on the symmetry of the MTSFM's modulation function
\begin{equation}  c_m = \left\{
\begin{array}{ll}
	\mathcal{J}_m^{1:K}\left(\{\alpha_k, -j\beta_k\}\right), & \varphi\left(t\right) \\

      \mathcal{J}_m^{1:K}\left(\{\alpha_k\}\right), & \varphi_e\left(t\right) \\
      
      \mathcal{I}_m^{1:K}\left(\{-j\beta_k\}\right), & \varphi_o\left(t\right) \\
\end{array} 
\right.
\label{eq:APP_I_9} 
\end{equation}

\section{Derivation of the MTSFM's AF}
\label{sec:AppendixIII}
Using the basebanded MTSFM time-series expression \eqref{eq:MTSFM_4} and the AF defined in \eqref{eq:AF}
\ifthenelse {\boolean{singleColumn}}
{\begin{IEEEeqnarray}{rCl}
\chi\left(\tau, \nu\right) = \dfrac{e^{-j\pi a_0\tau}}{T}\sum_{m,n}c_mc_n^*e^{-j\frac{\pi\left(m+n\right)\tau}{T}}\int_{-\infty}^{\infty}\rect\left(\dfrac{t-\tau/2}{T}\right)\rect\left(\dfrac{t+\tau/2}{T}\right)e^{j2\pi A t}dt
\label{eq:AF_1}
\end{IEEEeqnarray}}
{\begin{multline}
\chi\left(\tau, \nu\right) = \dfrac{e^{-j\pi a_0\tau}}{T}\sum_{m,n}c_mc_n^*e^{-j\frac{\pi\left(m+n\right)\tau}{T}} \times \\ \int_{-\infty}^{\infty}\rect\left(\dfrac{t-\tau/2}{T}\right)\rect\left(\dfrac{t+\tau/2}{T}\right)e^{j2\pi A t}dt
\label{eq:AF_1}
\end{multline}}
where $A = \left[\nu + \frac{\left(m-n\right)}{T}\right]$ and $c_m$ and $c_n^*$ represent $K$-dimensional GBFs.  The rectangular window functions establish the limits of integration $|t| \leq \left(\frac{T-|\tau|}{2}\right)$.  The expression in \eqref{eq:AF_1} then simplifies to 
\begin{IEEEeqnarray}{rCl}
\chi\left(\tau, \nu\right) = \dfrac{1}{T}\sum_{m,n}c_mc_n^*e^{-j\frac{\pi\left(m+n\right)\tau}{T}}\int_{-\frac{T-|\tau|}{2}}^{\frac{T-|\tau|}{2}}e^{j2\pi A t}dt
\label{eq:AF_2}
\end{IEEEeqnarray}
The integral in \eqref{eq:AF_2} evaluates to 
\begin{IEEEeqnarray}{rCl}
\left(T-|\tau|\right)\sinc\left[\pi\left(T-|\tau|\right)\left(\nu + \dfrac{\left(m-n\right)}{T}\right)\right].
\label{eq:AF_3}
\end{IEEEeqnarray}
Inserting this expression back in to \eqref{eq:AF_2} results in the final expression for the AF of the MTSFM waveform
\ifthenelse {\boolean{singleColumn}}
{\begin{multline}
\chi\left(\tau, \nu\right) = \left(\dfrac{T-|\tau|}{T}\right)\sum_{m,n}\mathcal{J}_m^{1:K}\left(\{\alpha_k, -\beta_k\}\right)\left(\mathcal{J}_n^{1:K}\left(\{\alpha_k, -\beta_k\}\right)\right)^*e^{-j\frac{\pi\left(m+n\right)\tau}{T}} \times \\ \sinc\left[\pi\left(\dfrac{T-|\tau|}{T}\right)\left(\nu T + \left(m-n\right)\right)\right].
\label{eq:AF_4}
\end{multline}}
{\begin{multline}
\chi\left(\tau, \nu\right) = \left(\dfrac{T-|\tau|}{T}\right) \times \\ \sum_{m,n}\mathcal{J}_m^{1:K}\left(\{\alpha_k, -\beta_k\}\right)\left(\mathcal{J}_n^{1:K}\left(\{\alpha_k, -\beta_k\}\right)\right)^* \times \\ e^{-j\frac{\pi\left(m+n\right)\tau}{T}} \sinc\left[\pi\left(\dfrac{T-|\tau|}{T}\right)\left(\nu T + \left(m-n\right)\right)\right].
\label{eq:AF_4}
\end{multline}}
The MTSFM's ACF directly follows from \eqref{eq:AF_4} by setting $\nu=0$.

\section*{Acknowledgment}
This work was funded by the internal investment program at the Naval Undersea Warfare Center Division Newport.  The author also wishes to acknowledge the reviewers of this article and their helpful suggestions throughout the review process. 
\ifCLASSOPTIONcaptionsoff
  \newpage
\fi


\bibliographystyle{IEEEtran}
\bibliography{Hague_MTSFM_IEEE_Trans_AES}

\begin{thebibliography}{10}
\providecommand{\url}[1]{#1}
\csname url@samestyle\endcsname
\providecommand{\newblock}{\relax}
\providecommand{\bibinfo}[2]{#2}
\providecommand{\BIBentrySTDinterwordspacing}{\spaceskip=0pt\relax}
\providecommand{\BIBentryALTinterwordstretchfactor}{4}
\providecommand{\BIBentryALTinterwordspacing}{\spaceskip=\fontdimen2\font plus
\BIBentryALTinterwordstretchfactor\fontdimen3\font minus
  \fontdimen4\font\relax}
\providecommand{\BIBforeignlanguage}[2]{{%
\expandafter\ifx\csname l@#1\endcsname\relax
\typeout{** WARNING: IEEEtran.bst: No hyphenation pattern has been}%
\typeout{** loaded for the language `#1'. Using the pattern for}%
\typeout{** the default language instead.}%
\else
\language=\csname l@#1\endcsname
\fi
#2}}
\providecommand{\BIBdecl}{\relax}
\BIBdecl

\bibitem{BluntIV}
S.~D. Blunt and E.~L. Mokole, ``Overview of radar waveform diversity,''
  \emph{IEEE Aerospace and Electronic Systems Magazine}, vol.~31, no.~11, pp.
  2--42, November 2016.

\bibitem{HaykinI}
S.~Haykin, ``Cognitive radar: a way of the future,'' \emph{IEEE Signal
  Processing Magazine}, vol.~23, no.~1, pp. 30--40, Jan 2006.

\bibitem{CognitiveI}
S.~Z. {Gurbuz}, H.~D. {Griffiths}, A.~{Charlish}, M.~{Rangaswamy}, M.~S.
  {Greco}, and K.~{Bell}, ``An overview of cognitive radar: Past, present, and
  future,'' \emph{IEEE Aerospace and Electronic Systems Magazine}, vol.~34,
  no.~12, pp. 6--18, Dec 2019.

\bibitem{Cook}
C.~Cook and M.~Bernfeld, \emph{Radar signals: an introduction to theory and
  application}, ser. Electrical science series.\hskip 1em plus 0.5em minus
  0.4em\relax Academic Press, 1967.

\bibitem{Rihaczek}
A.~Rihaczek, \emph{Principles of high-resolution radar}.\hskip 1em plus 0.5em
  minus 0.4em\relax McGraw-Hill, 1969.

\bibitem{Levanon}
E.~M. N.~Levanon, \emph{Radar Signals}.\hskip 1em plus 0.5em minus 0.4em\relax
  Wiley-Interscience, 2004.

\bibitem{Woodward}
P.~Woodward, \emph{Probability and Information Theory, with Applications to
  Radar}, ser. Radar Library.\hskip 1em plus 0.5em minus 0.4em\relax Artech
  House, 1980.

\bibitem{Wilcox}
C.~W. Wilcox, ``The synthesis problem for radar ambiguity functions,''
  \emph{Math. Res. Center, U.S. Army, Univ. of Wisconsin Rept. 157}, April
  1960.

\bibitem{Sussman}
S.~Sussman, ``Least-square synthesis of radar ambiguity functions,''
  \emph{Information Theory, IRE Transactions on}, vol.~8, no.~3, pp. 246--254,
  April 1962.

\bibitem{GladkovaI}
\BIBentryALTinterwordspacing
I.~Gladkova and D.~Chebanov, ``On a new extension of wilcox's method,'' in
  \emph{Proceedings of the 5th WSEAS International Conference on Applied
  Mathematics}, ser. Math'04.\hskip 1em plus 0.5em minus 0.4em\relax Stevens
  Point, Wisconsin, USA: World Scientific and Engineering Academy and Society
  (WSEAS), 2004, pp. 31:1--31:6. [Online]. Available:
  \url{http://dl.acm.org/citation.cfm?id=1378446.1378477}
\BIBentrySTDinterwordspacing

\bibitem{GladkovaII}
------, ``On the synthesis problem for a waveform having a nearly ideal
  ambiguity functions,'' in \emph{International Conference on Radar Systems},
  2004.

\bibitem{JianLiBookII}
H.~He, J.~Li, and P.~Stoica, \emph{Waveform design for active sensing systems:
  a computational approach}.\hskip 1em plus 0.5em minus 0.4em\relax Cambridge
  University Press, 2012.

\bibitem{jianLiII}
P.~{Stoica}, J.~{Li}, and Y.~{Xie}, ``On probing signal design for mimo
  radar,'' \emph{IEEE Transactions on Signal Processing}, vol.~55, no.~8, pp.
  4151--4161, Aug 2007.

\bibitem{PalomarIII}
L.~{Wu}, P.~{Babu}, and D.~P. {Palomar}, ``Transmit waveform/receive filter
  design for mimo radar with multiple waveform constraints,'' \emph{IEEE
  Transactions on Signal Processing}, vol.~66, no.~6, pp. 1526--1540, March
  2018.

\bibitem{RangaswamyI}
G.~{Cui}, H.~{Li}, and M.~{Rangaswamy}, ``Mimo radar waveform design with
  constant modulus and similarity constraints,'' \emph{IEEE Transactions on
  Signal Processing}, vol.~62, no.~2, pp. 343--353, Jan 2014.

\bibitem{AubryII}
A.~{Aubry}, A.~{DeMaio}, A.~{Farina}, and M.~{Wicks}, ``Knowledge-aided
  (potentially cognitive) transmit signal and receive filter design in
  signal-dependent clutter,'' \emph{IEEE Transactions on Aerospace and
  Electronic Systems}, vol.~49, no.~1, pp. 93--117, Jan 2013.

\bibitem{AubryIII}
A.~{Aubry}, A.~{De Maio}, M.~{Piezzo}, M.~M. {Naghsh}, M.~{Soltanalian}, and
  P.~{Stoica}, ``Cognitive radar waveform design for spectral coexistence in
  signal-dependent interference,'' in \emph{2014 IEEE Radar Conference}, May
  2014, pp. 0474--0478.

\bibitem{PrabhuBabuII}
L.~{Wu}, P.~{Babu}, and D.~P. {Palomar}, ``Cognitive radar-based sequence
  design via sinr maximization,'' \emph{IEEE Transactions on Signal
  Processing}, vol.~65, no.~3, pp. 779--793, Feb 2017.

\bibitem{AubryI}
A.~{Aubry}, A.~{De Maio}, B.~{Jiang}, and S.~{Zhang}, ``Ambiguity function
  shaping for cognitive radar via complex quartic optimization,'' \emph{IEEE
  Transactions on Signal Processing}, vol.~61, no.~22, pp. 5603--5619, Nov
  2013.

\bibitem{PrabhuBabuIII}
J.~{Song}, P.~{Babu}, and D.~P. {Palomar}, ``Sequence set design with good
  correlation properties via majorization-minimization,'' \emph{IEEE
  Transactions on Signal Processing}, vol.~64, no.~11, pp. 2866--2879, June
  2016.

\bibitem{PalomarI}
R.~{Zhou}, Z.~{Zhao}, and D.~P. {Palomar}, ``Unified framework for minimax mimo
  transmit beampattern matching under waveform constraints,'' in \emph{ICASSP
  2019 - 2019 IEEE International Conference on Acoustics, Speech and Signal
  Processing (ICASSP)}, May 2019, pp. 4150--4154.

\bibitem{PalomarII}
L.~{Wu} and D.~P. {Palomar}, ``Sequence design for spectral shaping via
  minimization of regularized spectral level ratio,'' \emph{IEEE Transactions
  on Signal Processing}, vol.~67, no.~18, pp. 4683--4695, Sep. 2019.

\bibitem{SoltanalianI}
A.~{Bose} and M.~{Soltanalian}, ``Constructing binary sequences with good
  correlation properties: An efficient analytical-computational interplay,''
  \emph{IEEE Transactions on Signal Processing}, vol.~66, no.~11, pp.
  2998--3007, June 2018.

\bibitem{HagueIV}
D.~A. Hague and J.~R. Buck, ``An experimental evaluation of the generalized
  sinusoidal frequency modulated waveform for active sonar systems,'' \emph{The
  Journal of the Acoustical Society of America}, vol. 145, no.~6, pp.
  3741--3755, 2019.

\bibitem{Blinchikoff}
J.~W. Taylor~Jr. and H.~J. Blinchikoff, ``Quadriphase code - a radar pulse
  compression signal with unique characteristics,'' \emph{IEEE Transactions on
  Aerospace and Electronic Systems}, vol.~24, no.~2, pp. 156--170, Mar 1988.

\bibitem{BluntI}
S.~D. Blunt, M.~Cook, J.~Jakabosky, J.~D. Graaf, and E.~Perrins,
  ``Polyphase-coded fm (pcfm) radar waveforms, part i: implementation,''
  \emph{IEEE Transactions on Aerospace and Electronic Systems}, vol.~50, no.~3,
  pp. 2218--2229, July 2014.

\bibitem{BluntIII}
P.~S. Tan, J.~Jakabosky, J.~M. Stiles, and S.~D. Blunt, ``On higher-order
  representations of polyphase-coded fm radar waveforms,'' in \emph{2015 IEEE
  Radar Conference (RadarCon)}, May 2015, pp. 0467--0472.

\bibitem{LevanonQuad}
N.~{Levanon} and A.~{Freedman}, ``Ambiguity function of quadriphase coded radar
  pulse,'' \emph{IEEE Transactions on Aerospace and Electronic Systems},
  vol.~25, no.~6, pp. 848--853, Nov 1989.

\bibitem{BluntII}
S.~D. Blunt, J.~Jakabosky, M.~Cook, J.~Stiles, S.~Seguin, and E.~L. Mokole,
  ``Polyphase-coded fm (pcfm) radar waveforms, part ii: optimization,''
  \emph{IEEE Transactions on Aerospace and Electronic Systems}, vol.~50, no.~3,
  pp. 2230--2241, July 2014.

\bibitem{Koivunen}
M.~{Bică} and V.~{Koivunen}, ``Generalized multicarrier radar: Models and
  performance,'' \emph{IEEE Transactions on Signal Processing}, vol.~64,
  no.~17, pp. 4389--4402, 2016.

\bibitem{CEOFDM}
S.~C. Thompson and J.~P. Stralka, ``Constant envelope ofdm for power-efficient
  radar and data communications,'' in \emph{2009 International Waveform
  Diversity and Design Conference}.\hskip 1em plus 0.5em minus 0.4em\relax
  IEEE, 2009, pp. 291--295.

\bibitem{Nehorai}
S.~{Satyabrata} and A.~{Nehorai}, ``Adaptive ofdm radar for target detection in
  multipath scenarios,'' \emph{IEEE Transactions on Signal Processing},
  vol.~59, no.~1, pp. 78--90, 2010.

\bibitem{Dattoli}
S.~Lorenzutta, G.~Maino, G.~Dattoli, M.~Richetta, A.~Torre, and C.~Chiccoli,
  ``{F}ourier expansions and multivariable {B}essel functions concerning
  radiation problems,'' \emph{Radiation Physics and Chemistry}, vol.~47, no.~2,
  pp. 183 -- 189, 1996.

\bibitem{Harris}
F.~Harris, ``On the use of windows for harmonic analysis with the discrete
  fourier transform,'' \emph{Proceedings of the IEEE}, vol.~66, no.~1, pp.
  51--83, Jan 1978.

\bibitem{Ricker}
D.~W. Ricker, \emph{Echo Signal Processing}.\hskip 1em plus 0.5em minus
  0.4em\relax Kluwer, 2003.

\bibitem{Lorenzutta}
S.~Lorenzutta, G.~Maino, G.~Dattoli, A.~Torre, and C.~Chiccoli,
  ``Infinite-variable bessel functions of the anger type and the fourier
  expansions,'' \emph{Reports on Mathematical Physics}, vol.~39, no.~2, pp. 163
  -- 176, 1997.

\bibitem{boyd}
J.~P. Boyd, \emph{Chebyshev and Fourier spectral methods}.\hskip 1em plus 0.5em
  minus 0.4em\relax Courier Corporation, 2001.

\bibitem{HagueDiss}
D.~A. Hague, ``The generalized sinusoidal frequency modulated waveform for
  active sonar,'' Ph.D. dissertation, Univ. of Massachusetts Dartmouth,
  Dartmouth, MA, 2015.

\bibitem{Auslander}
L.~Auslander and R.~Tolimieri, ``Characterizing the radar ambiguity
  functions,'' \emph{IEEE Transactions on Information Theory}, vol.~30, no.~6,
  pp. 832--836, November 1984.

\bibitem{HagueII}
D.~A. Hague, ``Transmit waveform design using multi-tone sinusoidal frequency
  modulation,'' in \emph{2017 IEEE Radar Conference (RadarConf)}, May 2017, pp.
  0356--0360.

\bibitem{HagueIII}
------, ``Optimal waveform design using multi-tone sinusoidal frequency
  modulation,'' in \emph{OCEANS 2017 - Anchorage}, September 2017, pp. 1--6.

\bibitem{Giacoletto}
L.~J. Giacoletto, ``Generalized theory of multitone amplitude and frequency
  modulation,'' \emph{Proceedings of the IRE}, vol.~35, no.~7, pp. 680--693,
  July 1947.

\bibitem{subBandMod}
E.~R. {Boecker}, ``Sub-band modulation in active sonar,'' PENNSYLVANIA STATE
  UNIV UNIVERSITY PARK APPLIED RESEARCH LAB, Tech. Rep., 2001.

\bibitem{NLFMI}
J.~A. Johnston and A.~C. Fairhead, ``Waveform design and doppler sensitivity
  analysis for nonlinear fm chirp pulses,'' \emph{Communications, Radar and
  Signal Processing, IEE Proceedings F}, vol. 133, no.~2, pp. 163--175, April
  1986.

\bibitem{HagueI}
D.~A. Hague and J.~R. Buck, ``The generalized sinusoidal frequency-modulated
  waveform for active sonar,'' \emph{IEEE Journal of Oceanic Engineering},
  vol.~PP, no.~99, pp. 1--15, 2016.

\bibitem{ultraSound_I}
T.~Misaridis and J.~A. Jensen, ``Use of modulated excitation signals in medical
  ultrasound. part i: Basic concepts and expected benefits,'' \emph{IEEE
  transactions on ultrasonics, ferroelectrics, and frequency control}, vol.~52,
  no.~2, pp. 177--191, 2005.

\bibitem{HagueV}
D.~A. Hague, ``Target resolution properties of the multi-tone sinusoidal
  frequency modulatedwaveform,'' in \emph{2018 IEEE Statistical Signal
  Processing Workshop (SSP)}.\hskip 1em plus 0.5em minus 0.4em\relax IEEE,
  2018, pp. 752--756.

\bibitem{Matlab}
``Matlab optimization toolbox,'' 2018b, the MathWorks, Natick, MA, USA.

\bibitem{Monga}
K.~{Alhujaili}, V.~{Monga}, and M.~{Rangaswamy}, ``Quartic gradient descent for
  tractable radar slow-time ambiguity function shaping,'' \emph{IEEE
  Transactions on Aerospace and Electronic Systems}, vol.~56, no.~2, pp.
  1474--1489, 2020.

\bibitem{Hofstetter}
R.~{Price} and E.~{Hofstetter}, ``Bounds on the volume and height distributions
  of the ambiguity function,'' \emph{IEEE Transactions on Information Theory},
  vol.~11, no.~2, pp. 207--214, April 1965.

\bibitem{Parker}
P.~Kuklinski and D.~A. Hague, ``Identities and properties of multi-dimensional
  generalized {B}essel functions,'' 2019.

\bibitem{jianLiIII}
P.~{Stoica}, H.~{He}, and J.~{Li}, ``New algorithms for designing unimodular
  sequences with good correlation properties,'' \emph{IEEE Transactions on
  Signal Processing}, vol.~57, no.~4, pp. 1415--1425, April 2009.

\bibitem{jianLiI}
H.~{He}, P.~{Stoica}, and J.~{Li}, ``Designing unimodular sequence sets with
  good correlations—including an application to mimo radar,'' \emph{IEEE
  Transactions on Signal Processing}, vol.~57, no.~11, pp. 4391--4405, Nov
  2009.

\bibitem{HagueVI}
D.~A. Hague, ``Nonlinear frequency moudlation using {F}ourier sine series,'' in
  \emph{2018 IEEE Radar Conference (RadarConf18)}, April 2018, pp. 1015--1020.

\bibitem{HagueVII}
D.~A. {Hague}, ``Generating waveform families using multi-tone sinusoidal
  frequency modulation,'' in \emph{2020 IEEE International Radar Conference
  (RADAR)}, 2020, pp. 946--951.

\bibitem{DattoliII}
G.~Dattoli and A.~Torre, \emph{Theory and Applications of Generalized Bessel
  Functions}.\hskip 1em plus 0.5em minus 0.4em\relax Aracne Editrice, 1996.

\end{thebibliography}

\ifCLASSOPTIONcaptionsoff
  \newpage
\fi



%\bibliographystyle{IEEEtran}
\begin{IEEEbiography}[{\includegraphics[width=1in,height=1.25in,clip,keepaspectratio]{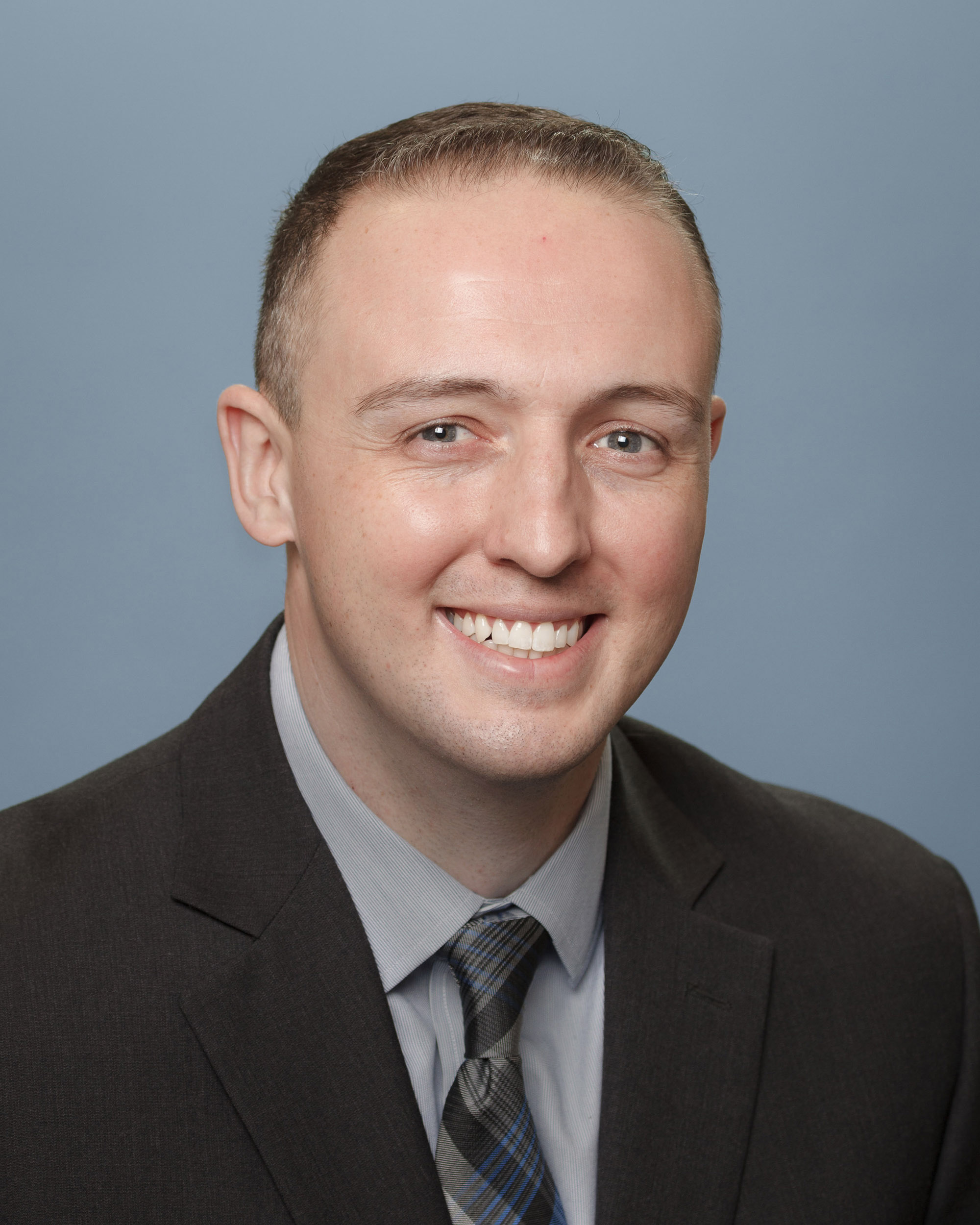}}]{David A. Hague}
%\begin{IEEEbiography}{David A. Hague}
received the B.S., M.S., and Ph.D. degrees in electrical engineering from the University of Massachusetts Dartmouth, Dartmouth, MA, USA, in 2005, 2012, and 2015, respectively.  He is a veteran of the U.S. Army and did tours of duty in Afghanistan in 2002 and Kosovo in 2006.  From the end of 2006 until August 2009, he worked for Raytheon Integrated Defense Systems, Tewksbury, MA, USA, where he developed signal processing software for a variety of radar systems.  He returned to UMass Dartmouth in the fall of 2009 as a Research Assistant at the Acoustic Signal Processing Laboratory run by Dr. John R. Buck to pursue graduate study with a concentration in acoustic signal processing.  Dr. Hague received the Science, Mathematics, and Research for Transformation (SMART) program scholarship in 2010 and 2011 to pursue his graduate studies with the Naval Undersea Warfare Center (NUWC) as his sponsoring facility.  David then joined NUWC as a full-time employee in September of 2015 conducting basic and applied research.  His main research interests are radar/sonar signal processing, time/frequency analysis, waveform design, compressive sensing, and graph signal processing.  He is a member of the Acoustical Society of America and the IEEE Signal Processing Society (SPS).  David has additionally served as chair of the IEEE SPS Providence chapter in 2018 and 2019 and has also served on the technical committee for the 2017 and 2019 IEEE Underwater Acoustic Signal Processing Workshop.
\end{IEEEbiography}

\end{document}